    \documentclass[twocolumn]{aastex701}
    \usepackage{hyperref}
    \usepackage{amsfonts}
    \usepackage{amsmath}
    \usepackage{upgreek}
    \usepackage{graphicx}
    \usepackage{url}
    \usepackage{hyperref}
    \usepackage{float}
    \usepackage{color}
    \usepackage[utf8]{inputenc}
    \newcommand{\brunt}[0]{Brunt-V\"ais\"al\"a}
    \newcommand*{\rd}[2]{\frac{\mathrm{d}#1}{\mathrm{d}#2}}
    \newcommand*{\pd}[2]{\frac{\partial#1}{\partial#2}}
    \newcommand*{\p}[1]{\left(#1\right)}
    \newcommand*{\s}[1]{\left[#1\right]}
    \newcommand*{\z}[1]{\left\{#1\right\}}

\defcitealias{mankovich2021diffuse}{MF21}
\defcitealias{sur2025_fuzzycores}{S25}

\begin{document}

\title{Saturn's Evolutionary History and Seismology: Survival of Deep Stably Stratified Regions in Evolutionary Models of Saturn Consistent with Ring Seismology}

\author[orcid=0000-0001-8283-3425]{Yubo Su}
\email{yubo.su@utoronto.ca}
\affiliation{Department of Astrophysical Sciences, Princeton University, Princeton, NJ 08544, USA}
\affiliation{Canadian Institute for Theoretical Astrophysics, University of Toronto, 60 St George Street, Toronto, M5S 3H8 Ontario, Canada}

\author[orcid=0000-0001-9420-5194]{Janosz W. Dewberry}
\email{jdewberry@umass.edu}
\affiliation{Department of Astronomy, University of Massachusetts Amherst, 710 N Pleasant St, Amherst, MA 01003, USA}

\author[orcid=0000-0001-6708-3427]{Roberto Tejada Arevalo}
\email{arevalo@princeton.edu}
\affiliation{Department of Astrophysical Sciences, Princeton University, Princeton, NJ 08544, USA}

\author[orcid=0000-0001-6635-5080]{Ankan Sur}
\email{ankansur@ucla.edu}
\affiliation{Department of Astrophysical Sciences, Princeton University, Princeton, NJ 08544, USA}
\affiliation{Department of Earth, Planetary, and Space Sciences, University of California Los Angeles, 595 Charles E Young Dr E, LA, CA 90095}

\author[orcid=0000-0002-3099-5024]{Adam Burrows}
\email{burrows@astro.princeton.edu}
\affiliation{Department of Astrophysical Sciences, Princeton University, Princeton, NJ 08544, USA}

\begin{abstract}

With recent advances in the modeling of the solar system giant planets, rapid
progress has been made in understanding the remaining questions pertaining to
their formation and evolution.
However, this progress has largely neglected the significant constraints on the
interior of Saturn's structure imposed by the observed oscillation frequencies
in its rings.
Here, we study initial conditions for Saturn's evolution that, after
$4.56\;\mathrm{Gyr}$ of evolution, give rise to planetary structures admitting
oscillation frequencies consistent with those observed via Saturn's ring
seismology.
Restricting our attention to models without compact rocky cores,
we achieve simultaneous good agreement with most observed properties of Saturn
at the level of current evolutionary models and with key frequencies in the
observed oscillation spectrum.
Our preliminary work suggests that Saturn's interior stably stratified region
may be moderately less extended ($\sim 0.4$--$0.5R_{\rm Sat}$) than previously
thought, which is important for reconciling the seismic constraints with
evolutionary models.
We also tentatively find that the deep helium gradients inferred by previous,
static structural modelling of Saturn's ring seismology may not be required to
reproduce the observed seismology data.

\end{abstract}


\section{Introduction}

In traditional models of planet formation, gas giant planets are thought to form
via a two-stage process, where the formation of a sufficiently massive rocky
core is followed by runaway gas accretion \citep[e.g.][]{pollack1996_coreacc}.
Such a process is expected to form planets with a simple two-zone structure
consisting of a compact rocky core and an extended gaseous envelope.
One straightforward test of this hypothesis comes from measurements of
Jupiter's zonal gravity field by \textit{Juno} \citep{bolton2017jupiter}.
However, these measurements yield just a few different spatially-weighted
combinations of Jupiter's density distribution that are predominantly sensitive
to the outer regions of the planet.
This difficulty causes significant variations in Jupiter's inferred interior
structure to arise depending on the adopted model for its near-surface zonal
winds \citep[see e.g.][]{wahl2017comparing, debras2019new, militzer2022juno,
miguel2022_jupstructure, howard2023, militzer2024_layermodels}.
Nevertheless, a consistent qualitative feature throughout these works is
evidence for extended composition gradients in Jupiter's deep interior, posing
new challenges for the formation of the Solar System gas giants \citep[see][for
excellent reviews]{stevenson2020jupiter, helled2022revelations}.
Efforts to place such ``fuzzy cores'' in the context of the gas giants'
formation and evolution typically invoke (1) partial erosion and dilution of an
initially compact core \citep{guillot2004_jup, moll2017_juperosion}; (2)
planetesimal collisions \citep{liu2019_dilute, meier2025_dilutecollision}; or
(3) the conclusion that they are a natural consequence of the mass accretion
process \citep{ormel2021_pebblegradient, stevenson2022_mixing,
bodenheimer2025_saturn}.
While such efforts are ongoing, it is clear that the inferred existence of fuzzy
cores today place stringent constraints on the initial conditions and subsequent
evolution of our giant planets' structures \citep{fuentes2023_erosion,
knierim2024_erosion, markham2024_rain, fuentes2025_energyerosion,
tejadaarevalo2025_jupiterfuzzy, sur2025_fuzzycores, knierim2025_constraints}.

Given the importance of accurately understanding giant planet fuzzy cores,
observational constraints on their properties are at a premium.
As such, it is a fortuitous coincidence that the other Solar System gas giant,
Saturn, has a striking ring system that doubles as an exquisite seismograph for
the planet's oscillation mode spectrum \citep{stevenson1982a,
hedman2013kronoseismology, Mankovich2020a, french2021_krono, afigbo2025_krono7}.
Unlike zonal gravity field measurements, oscillation mode frequencies can give
direct insights into the structure of Saturn's deep interior
\citep{fuller2014saturn, mankovich2019cassini, mankovich2021diffuse}.
Note that, while there are tentative detections of Jupiter's normal mode
oscillations from ground-based interferometry (\citealp{gaulme2011detection,
gaulme2015review}, though see \citealp{shaw2022_pmode} who find
no evidence of p-mode power down to a sensitivity of $\sim 16\;\mathrm{cm/s}$)
and from a detailed analysis of its gravitational field from \emph{Juno} data
\citep{durante2022juno}, no data with the constraining power of the measurements
made for Saturn are available at this time\footnote{
Note that the oscillations putatively detected by \citet{gaulme2011detection}
have surface radial velocities of $\sim50\;\mathrm{cm/s}$, while the typical
surface velocities probed by ring seismology are $\sim 0.05\;\mathrm{cm/s}$
\citep{fuller2014saturn}!
While these two precisions are not directly comparable due to the differing
frequency regimes of p- and $g$-mode oscillations, they nevertheless highlight the
unparallel precision afforded by ring seismology.
}.
Nevertheless, it is clear that the incorporation of seismic constraints into
giant planet evolution models are an important step towards
understanding their deep interior structures.

In their seminal work, \citet{mankovich2021diffuse}
\citepalias[hereafter][]{mankovich2021diffuse} found that the combination of
Saturn's seismology and gravity data provided strong evidence for large-scale
stable stratification extending from the core of the planet to $\sim 0.6R_{\rm
Sat}$ at the present day, attributed to compositional gradients in both helium
and heavier elements.
Subsequent work has provided additional inferences from the numerous oscillation
modes identified \citep{dewberry2021_satrings, dewberry2022_saturnwinds,
mankovich2023_satrot2}, though these have largely advanced the characterization
of Saturn's structure near its surface. In the opposite direction, modern
evolutionary models find that the stably stratified interiors of giant planets
become smaller over time in both
1D planetary evolution calculations \citep{knierim2024_erosion,
tejadaarevalo2025_jupiterfuzzy, sur2025_fuzzycores} and 3D hydrodynamical simulations
\citep{fuentes2023_erosion, hindman2023_rotationconvgrowth,
fuentes2024_rotatingsemiconv, zhang2025_fuentes_hydrosimentrainment}
(see Section~\ref{ss:core_erosion} for further discussion).
The apparent conflicting demands of the present-day seismology and evolutionary
perspectives pose a quantitative question: What initial conditions are required
for a primordially stratified composition in Saturn to survive to the present
day with a radial extent consistent with seismological constraints?

In this paper, we use the planet evolution code \texttt{APPLE}
\citep{sur2024_apple} to explore evolutionary histories for Saturn that are
consistent with the seismological constraints from its rings.
We seek two classes of answers to the question posed above.
First, we aim to identify models that recover the structures inferred by
\citetalias{mankovich2021diffuse} by imposing a primordial gradient in the
fractional helium abundance (relative to hydrogen, i.e., $Y' \equiv Y / (1 - Z)$
where $Y$ and $Z$ are the absolute helium and heavy element mass fractions).
Second, we explore models that begin with a uniform value of $Y'$ set by the
protosolar value, which has been the standard assumption for evolutionary models
(e.g., \citealp{stevenson1977_rain, FortneyHubbard2003, Nettelmann2015,
Pustow2016, mankovich2020_ognp, sur2025_fuzzycores}).
Through this exploration, we aim to isolate the constraints that motivate the
proposed $Y'$ gradient and quantify its effect in reproducing the observational
constraints.
We describe our methods in Section~\ref{s:methods}, present our models and their
seismology in Section~\ref{s:results}, discuss our results in
Section~\ref{s:disc}, and conclude in Section~\ref{s:summary}.

\section{Methods}\label{s:methods}

In this section, we describe the methods we will use to evaluate our
models.
We first describe the observations that we seek to reproduce, and then we
describe the numerical methods that we use to perform our planet modeling.

\subsection{Observations}

To evaluate our evolutionary models, we seek to match Saturn's
observed equatorial radius $R_{\rm eq}$, surface effective temperature $T_{\rm
eff}$, second $J_2$ and fourth $J_4$ gravitational zonal harmonics, and bulk rotation
period $P_{\rm Sat}$\footnote{
While many other works attempt to fit for the surface $Y$ and $Z$ abundances
\citep{koskinen2018_satyatm, guillot2023_satz}, we defer this for future work:
as we will discuss later, we do not include rocky cores in our analysis
for technical reasons, so the $Z$ distributions we obtain will
likely differ significantly from better-fitting models that include compact, rocky cores.
}.
These values are provided in Table~\ref{tab:params}.
\begin{table}
    \begin{tabular}{|l l l|}
    \hline
    Quantity & Value & Reference\\
    \hline
        $R_{\rm eq}$ &
        $6.0268 \times 10^{10}\;\mathrm{cm}$ &
        (1)
        \\
        $T_{\rm eff}$ &
        $96.67$ K &
        (2)
        \\
        $J_2$ [$\times 10^6$] &
        $16290.573$ &
        (3)
        \\
        $J_4$ [$\times 10^6$] &
        $-935.314$ &
        (3)
        \\
        $P_{\rm Sat}$ &
        $38082\;\mathrm{s}$ &
        (4)
        \\\hline
        $\omega_{\rm Maxwell} / \Omega_{\rm dyn}$
        &
        $1.6308$
        &
        (5)
        \\
        $\omega_{\rm W87.19} / \Omega_{\rm dyn}$
        &
        $1.6403$
        &
        (5)
        \\
        $\omega_{\rm W84.64} / \Omega_{\rm dyn}$ &
        $1.7153$ &
        (5)
        \\
        $\omega_{\rm W76.44} / \Omega_{\rm dyn}$ &
        $1.9996$ &
        (5)
        \\
        $A_{\rm W84.64} / A_{\rm 76.44}$ &
        $3.80 \pm 0.40$ &
        (6)
        \\
        $A_{\rm W84.64} / A_{\rm 87.19}$
        &
        $4.55 \pm 0.20$
        &
        (6)
        \\
        \hline
    \end{tabular}
    \caption{
    Structural parameters (top) and seismological ones (bottom).
    Oscillation frequencies are given in the inertial frame.
    For reference, note that $2\pi / (\Omega_{\rm dyn}P_{\rm Sat}) \simeq 0.38$.
    References are: (1) \citet{seidelmann2007_satrad}, (2)
    \citet{li2012_satTeff}, (3) \citet{iess2019measurement},
    (4) \citet{mankovich2023_satrot2},
    (5) \citet{french2019_krono3}, (6) \citet{afigbo2025_krono7}.
    }\label{tab:params}
\end{table}

A few tens of oscillation frequencies have been identified
in the rings \citep{hedman2013kronoseismology, hedman2014krono2,
french2019_krono3, hedman2019_krono4, french2021_krono}, and recent works have
used all of these frequencies to infer very tight constraints on the bulk
rotation rate and near-surface rotation profile \citep{mankovich2019cassini,
mankovich2023_satrot2}.
However, most of these oscillation modes correspond to high angular number $f$
modes that are largely insensitive to the deep interior structure of the planet.
Instead, the four frequencies attributed to $\ell=2$ oscillation modes are the
most constraining for the deep internal structure of Saturn
\citep[though the $m=3$ triplet also carries significant constraining power,
see][]{dewberry2021_satrings}.
These are the Maxwell, W87.19, W84.64, and W76.44 frequencies \citep[following
the naming conventions of][]{colwell2009_ringnames, french2019_krono3}.
Among these, the closely-spaced Maxwell and W87.19 frequencies are quite close
to the inertial wave frequency range, which occurs when the wave frequency in
the co-rotating frame of the planet $\sigma$ satisfies $|\sigma| < 2\Omega_{\rm
Sat}$, where $\Omega_{\rm Sat} = 2\pi / P_{\rm Sat}$ is Saturn's rotational
angular frequency.
The remaining W84.64 and, more importantly, the W76.44 frequencies furnish the
strongest evidence for Saturn's deeply stratified core \citep{fuller2014saturn,
mankovich2021diffuse}.
The frequencies of all four of these ring features are in
Table~\ref{tab:params}, where they are given relative to Saturn's equatorial
dynamical frequency $\Omega_{\rm dyn} \equiv \sqrt{G M_{\rm Sat} / R_{\rm
eq}^3}$, a characteristic frequency scale for gravitationally-restored $f$
modes.
Note that \citet{afigbo2025_krono7} performed fits to the optical depth
variations of the density rings in the waves to constrain the amplitudes of the
planetary oscillations sourcing these variations, and their results are in
Table~\ref{tab:params}.
However, these amplitudes are proportional to the surface gravitational
perturbations of their corresponding planetary mode, so they reflect the product
of the mode amplitude and the mode eigenfunction's surface gravitational
perturbation. Since calculating the oscillation modes of the planet yields only
the latter quantity, the mode amplitude is unknowable without further
assumptions.
One common assumption, which we will adopt here, is energy equipartition among
modes \citep[e.g.,][]{fuller2014saturn, dewberry2022_saturnwinds}.
Under this assumption, the ratios of mode eigenfunctions' surface gravity
perturbations can be directly compared to the observed ratio of optical depth
variations.
This assumption is of course unlikely to be exact, and so agreement with the
observed amplitude ratio should be evaluated only at a qualitative level.

\subsection{Numerics}\label{ss:methods}

We solve for Saturn's evolution using the planetary evolution code
\texttt{APPLE} \citep{sur2024_apple}.
Note that, in addition to the thermal, compositional, and structural evolution
of the planet, \texttt{APPLE} solves for its deformation as well, using the
fourth-order theory of figures \citep{Nettelmann2017}, and includes an
angle-averaged rotational contribution to the equation of hydrostatic
equilibrium \citep{sur2024_apple}.
The evolution of the planet's rotation rate is governed by the conservation of angular momentum.
We adopt the hydrogen-helium equation of state from \citet{chabrier2021new}, and
we treat the heavy elements using the AQUA equation of state
\citep{haldeman2020_aqua}, unless otherwise specified (see
Section~\ref{ss:models2}).
The AQUA EOS contains an error in the entropy in \citet{Mazevet2019},
and was corrected in \citet{Mazevet2021}.
This error does not affect the density and has now been shown to be
inconsequential for the evolution of gas giants \citep{sur2026_exoplanet}
and even for Uranus and Neptune \citep{Tejada2025_UN}.
Given that this correction does not affect the density, it is not expected to affect the results of this paper.
Finally, note that \texttt{APPLE} models semiconvective energy transport
via the $R_\rho$ parameterization.
While this parameterization is often used in planetary evolution
studies \citep[e.g.,][]{mankovich2016_rrho, sur2024_apple},
it does not model physical semiconvection \citep[e.g.,][]{
leconte2012_doublediff, wood2013_ddconv,
sur2025_rrho}.
For clarity, we do not consider semiconvective energy transport in
the models presented in this paper, so we set $R_\rho = 1$\footnote{
In preliminary tests, we found that near-unity $R_\rho$ do not substantially
change our results, while smaller values of $R_\rho$ result in more convective
mixing \citep{sur2025_rrho} and even frequent numerical instability
(Appendix~\ref{app:stairsteps}).
Given the lack of physical motivation for the $R_\rho$ convective transport
prescription, we neglect it for the purposes of this paper.
}.

The atmospheric boundary conditions are adopted from \citet{chen2023_bcs}, and
we follow \citet{sur2025_fuzzycores} in interpolating these boundary conditions
as functions of the helium and heavy element mass fractions at the base of the
atmosphere.
We model hydrogen-helium immiscibility \citep{stevenson1977_rain} using the
``scheme B'' prescription\footnote{
The scheme A miscibility scheme in \texttt{APPLE} \citep{sur2024_apple}
identifies miscibility regions in pressure space and imposes a constant
degree of helium advection in immiscible regions.
This scheme much less numerically stable than scheme B due to a discontinuous
advection strength as $Y$ crosses the immiscibility threshold and due to the
sensitivity of immiscibility region identification when the structure is nearly
tangent to the miscibility curve.
}
described in \citet{sur2024_apple} (as also adopted in
\citealp{helled2025_mespa}).
In this scheme, the strength of helium advection in immiscible regions is
proportional to the excess helium fraction beyond miscibility ($\propto Y -
Y_{\rm misc}$) and drives the helium concentration gradient with a
characteristic strength $\propto (Y - Y_{\rm misc}) / \mathcal{H}_{\rm r}$;
the scale parameter $\mathcal{H}_{\rm r}$ sets the characteristic scale height
of helium demixing, and we take its value to be $2 \times 10^8\;\mathrm{cm}$.
We use the miscibility curves of \citet{lorenzen09, lorenzen11}.
Since improvements to the understanding of hydrogen-helium miscibility are
ongoing \citep[e.g.][]{brygoo2021_misc}, we adopt the usual practice of using a
temperature shift to the miscibility curves such that the resulting surface
helium abundances match atmospheric observations
\citep[e.g.,][]{mankovich2016_rrho, tejadaarevalo2025_jupiterfuzzy,
sur2025_fuzzycores}.
Throughout this work, we adopt $\Delta T = +410\;\mathrm{K}$.
Note that, since the miscibility curves dynamically adjust to the pressure and
temperature profiles of the planet, the resulting evolution can easily be stiff,
with difficulty achieving converged timesteps due to the simultaneously
adjusting structure, composition flux, and miscibility curves.
To avoid this, we adopt a time lag prescription for the local critical
immiscibility threshold $Y_{\rm m}'(P, T)$ (above which helium demixes), such
that $Y_{\rm m}'$ approaches the instantaneous value derived from the
miscibility curve on a timescale $t_{\rm Y_m, lag}$.
Since this is a purely numerical prescription, smaller values of $t_{\rm Y_m,
lag}$ are preferred, and we adopt $t_{\rm Y_m, lag} = 5\;\mathrm{Myr}$
throughout this work, one half our maximum timestep.
We confirmed that different choices of $t_{\rm Y_m, lag} \gtrsim
1\;\mathrm{Myr}$ do not affect the observed properties of the planet.

To calculate the oscillation frequencies of our Saturn models, we use the
partial differential solver described in \citet{dewberry2021_satrings,
dewberry2022_saturnwinds}.
We refer the reader to these papers for further details on the numerical methods
of this approach.
In brief, the code takes in a smoothed\footnote{
All fields are interpolated using a cubic spline except for the \brunt{}
frequency, which is linearly interpolated to preserve non-negativity.
}
thermodynamic profile of the planet and its deformation via the shape functions
from the theory of figures.
Then it uses a 2D pseudo-spectral method to compute linearized oscillation
modes, treating centrifugal flattening and the Coriolis force
non-perturbatively.
We use $120$ Gauss-Lobatto collocation grid points in the (quasi-)radial
direction, and project onto a spectral expansion in spherical harmonics
truncated at degree $\ell = |m|+30$ (the projection integrals over planetary
latitude are computed with a $257$-point Gauss-Legendre quadrature).
We find that increasing radial and spectral resolutions affects mode frequencies
and surface gravitational perturbations negligibly (though see
Section~\ref{ss:models0}), except for oscillations with frequencies
$\leq2\Omega_{\rm Sat}$ in the rotating frame.
For these modes, which generally include the $n=2$ $g$-mode, the ill-posed
nature of the inviscid eigenvalue problem for inertial waves in a rotating shell
\citep[e.g.,][]{ogilvie2009hyperbolic} necessitates the inclusion of frictional
terms.
We use a constant kinematic viscosity $\nu/(R^2\Omega_{\rm dyn})=10^{-5},$ and
truncate at $\ell=|m|+50$ to compute modes in this frequency regime.
The relatively low spectral resolution needed to resolve our modes is consistent
with the fact that they are low-wavenumber oscillations with large frequency
separations.

We perform searches for eigenvalues of the oscillation equations (corresponding
to oscillation modes) in the range $\omega \in [1.5, 2.4] \Omega_{\rm dyn}$,
though the specific frequency grid used for the search was varied among models
and timesteps in order to reliably locate the low-order modes of
interest\footnote{
Specifically, we first identified all relevant mode frequencies with a low-resolution sweep, and show the results of narrow-band high-spectral-resolution searches.
As such, the absence of modes in many frequency bands does not denote a true absence of oscillation modes.
}

Mode eigenfunctions are normalized following the convention of
\citet{dahlentromp1999} \citep[cf.][]{dewberry2021_satrings} such that the
Lagrangian displacements are orthonormal.
Then, mode energies in the co-rotating frame are given by $\varepsilon_i =
2\sigma_i^2$ \citep{Schenk2001}, so we divide the mode amplitudes under
normalization by $\sigma_i / \sigma_{\rm W84.64}$ to ensure energy equipartition
\citep{fuller2014saturn, mankovich2021diffuse}.
The modes are then ordered by the $\ell=2$ component of their surface
gravitational potential perturbation, denoted $\delta \Phi_{\ell=2,
\mathrm{surf}}$, which are thought to be predominantly responsible for exciting
the $m=2$ surface density fluctuations measured in the rings.

\section{Results}\label{s:results}

In this section, we aim to reproduce interior structure profiles as
consistent as possible with the combined physical constraints on Saturn's bulk
properties considered in previous works
\citep[e.g.][]{sur2025_fuzzycores, sur2025_rrho}
and those from seismology \citep{hedman2013kronoseismology,
mankovich2021diffuse}.
We will consider four models in this section, and their attributes are summarized in Table~\ref{tab:models}.
\begin{table*}
    \begin{tabular}{l|l l l l l}
         & Section & Initial$Y'$ Gradient & Metal EOS & Fits W84.64/W76.44? & Comments\\\hline
         Model 1 & \ref{ss:models0} & No & AQUA & No & Coreless ver.\ of \citetalias{sur2025_fuzzycores}\\
         Model 2 & \ref{ss:models1} & Yes & AQUA & Yes & Resembling \citetalias{mankovich2021diffuse}\\
         Model 3 & \ref{ss:models2} & No & AQUA & $\sim 5\%$ & \\
         Model 4 &  \ref{ss:models2} & No & Postperovskite & $\sim 5\%$ & \\
    \end{tabular}
    \caption{
    Summary of models considered in Section~\ref{s:results} and their
    differences.
    \citetalias{sur2025_fuzzycores} refers to \citet{sur2025_fuzzycores}.
    We focus our attention on the W84.64/W76.44 mode for the purposes of this
    table, as they provide the leading-order constraints on the shape of the
    stably stratified region.
    None of the models shown here reproduce all four known $m=2$ modes, as they
    struggle with the W87.19/Maxwell doublet.
    }
    \label{tab:models}
\end{table*}

\subsection{Comparison with Previous \texttt{APPLE} Models}\label{ss:models0}

\begin{figure*}
    \centering
    \includegraphics[width=\textwidth]{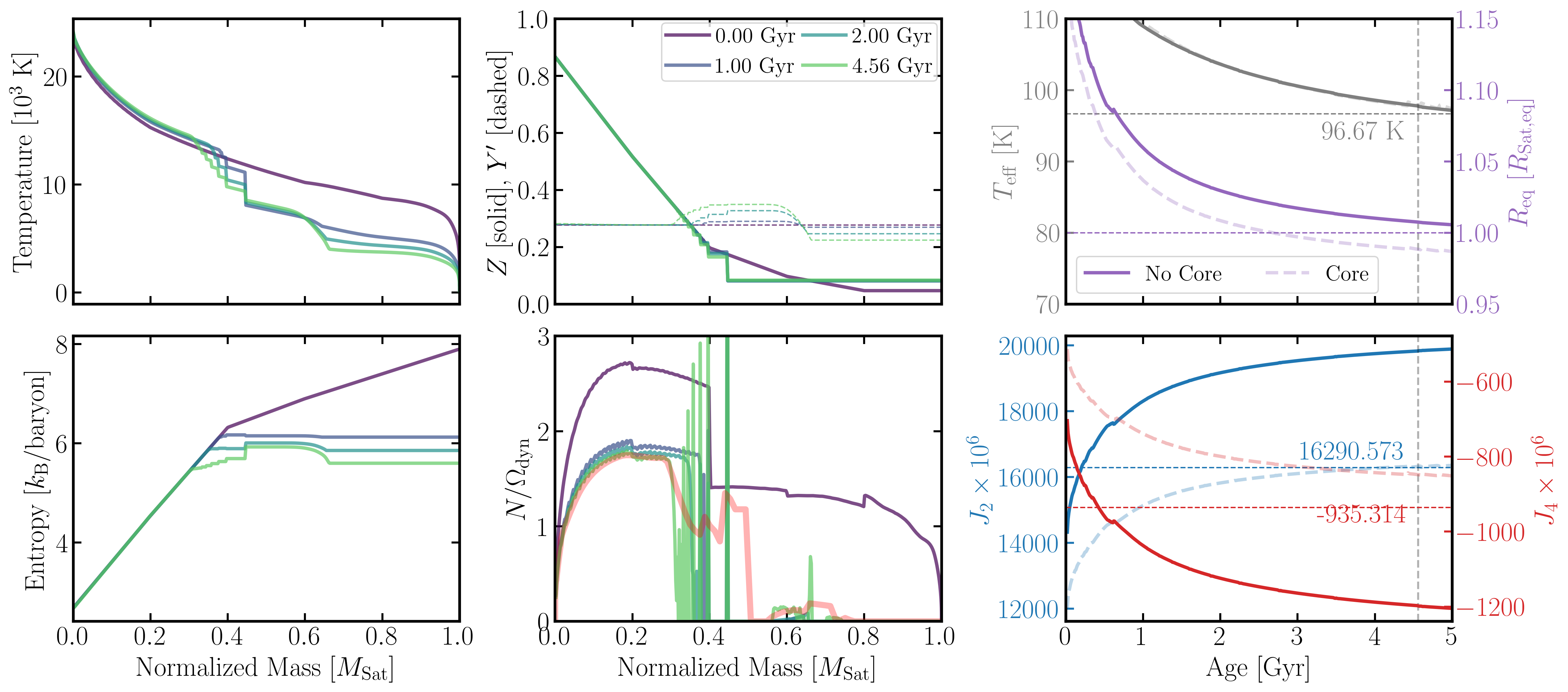}
    \caption{
    Evolution of the best-fitting Saturn model from \citet{sur2025_fuzzycores}
    with its $4M_{\oplus}$ compact core removed and restributed uniformly throughout the
    envelope.
    The left four panels show the profiles of the internal temperature, entropy,
    heavy element mass fraction ($Z$), fractional helium mass fraction ($Y'
    \equiv Y / (1 - Z)$), and the \brunt{} frequency $N$ normalized to $\Omega_{\rm
    dyn} \equiv \sqrt{G M_{\rm Sat} / R_{\rm eq}^3}$ as functions of the
    normalized mass coordinate of Saturn at four times (legend in upper middle
    panel).
    The large variations in $N$ are due to the discontinuities in the
    composition and thermal profiles of the planet (``stair steps''), which are
    commonly seen in 1D planetary evolution models
    \citep[e.g.][]{vazan2018jupiter, tejadaarevalo2025_jupiterfuzzy}.
    The smoothed $N$ profile that is used to compute oscillation
    modes is shown as the faint, thick red line.
    The right two panels show the temporal evolution of the effective
    temperature $T_{\rm eff}$, the equatorial radius (in units of the measured
    equatorial radius of Saturn given in Table~\ref{tab:params}), and the
    gravitational zonal harmonics $J_2$ and $J_4$.
    The vertical grey dashed line in the right two panels shows $t =
    4.56\;\mathrm{Gyr}$, the current age of Saturn.
    The final rotation rate is $1.01 P_{\rm Sat}$.
    For comparison, the faint dashed lines in the right two panels show the evolution
    for the original model from \citet{sur2025_fuzzycores} with a rocky core.
    We see that the presence of the rocky core reduces $J_2$ and $J_4$ by $\sim
    20$--$30\%$, while the effect on $T_{\rm eff}$ and $R_{\rm eq}$ is much
    smaller.
    }\label{fig:sat_rotnocore}
\end{figure*}

Note that, as of this work, our oscillation frequency solver does not support a
compact rocky core, which requires separate treatment due to the large density
discontinuity at the interface.
Such cores are present in the best-fitting evolutionary Saturn models obtained
with \texttt{APPLE} \citep[e.g.][]{sur2025_fuzzycores}.
As such, as a point of comparison for our later results, we first present the
seismology results for a model similar to the best-fitting Saturn model from
\citet{sur2025_fuzzycores}, but we remove the $4M_{\oplus}$ core and distribute
its heavy element content uniformly throughout the envelope.
We also adjust the initial rotation rate such that the final rotation rate
(under angular momentum conservation) remains consistent with the observed
values.
The resulting evolutionary track is shown in Fig.~\ref{fig:sat_rotnocore}, and a
comparison to the published model is shown in the light dashed lines in the two
right panels.
It can be seen that the removal of the core results in much larger magnitudes
for $J_2$ and $J_4$, a natural result of the lessened degree of central mass
concentration in the planet.
For reference, note that our models obtain a final rotation rate of $1.01P_{\rm
sat}$, where $P_{\rm sat}$ is taken from \citet{mankovich2023_satrot2}; the
typical range of estimates in the literature from static models is around
$3\;\mathrm{min} \approx 0.005P_{\rm sat}$ of this value \citep{read2009_satrot,
helled2015_satrot, mankovich2019cassini}.

\begin{figure}
    \centering
    \includegraphics[width=0.9\columnwidth]{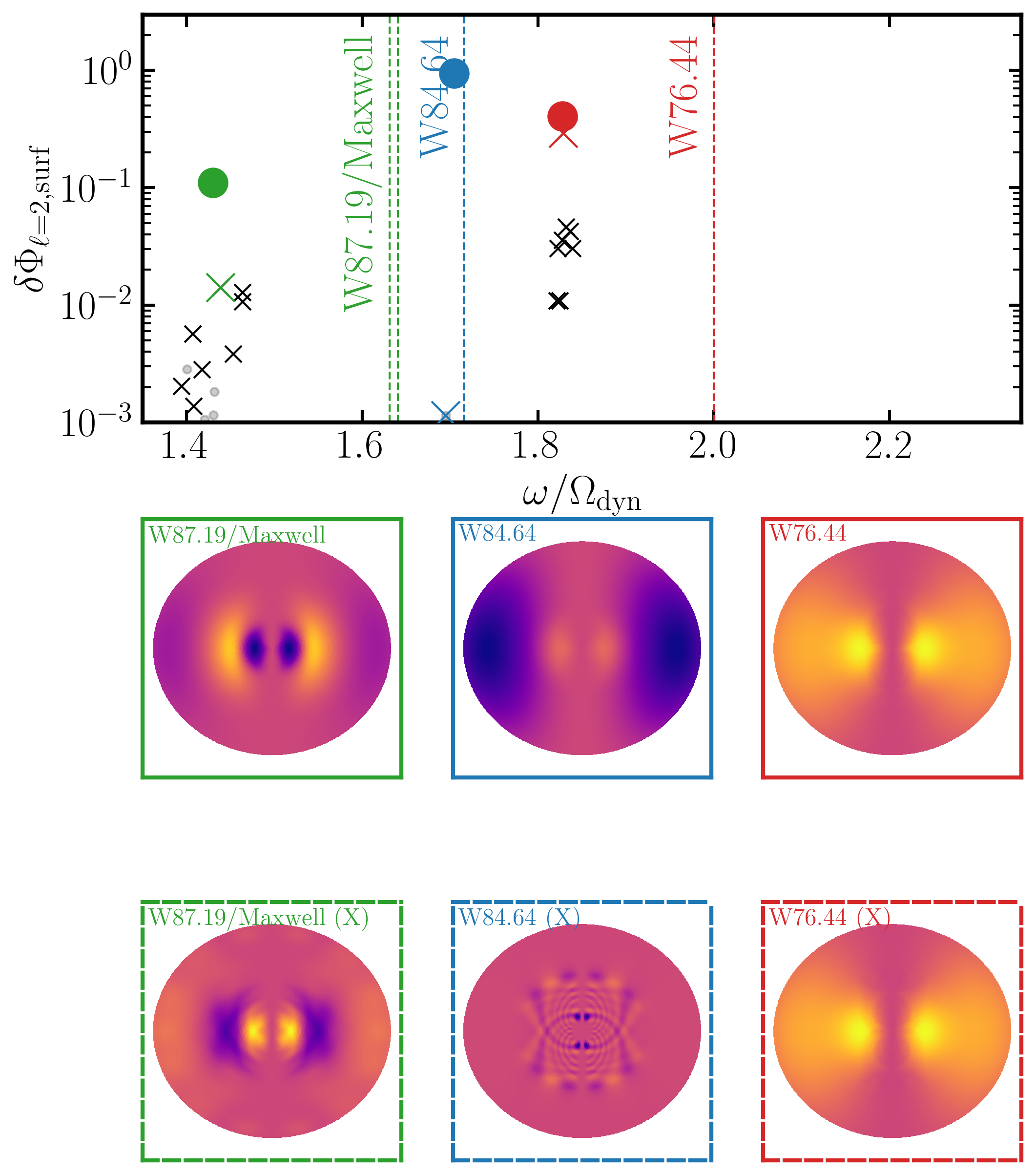}
    \includegraphics[width=0.9\columnwidth]{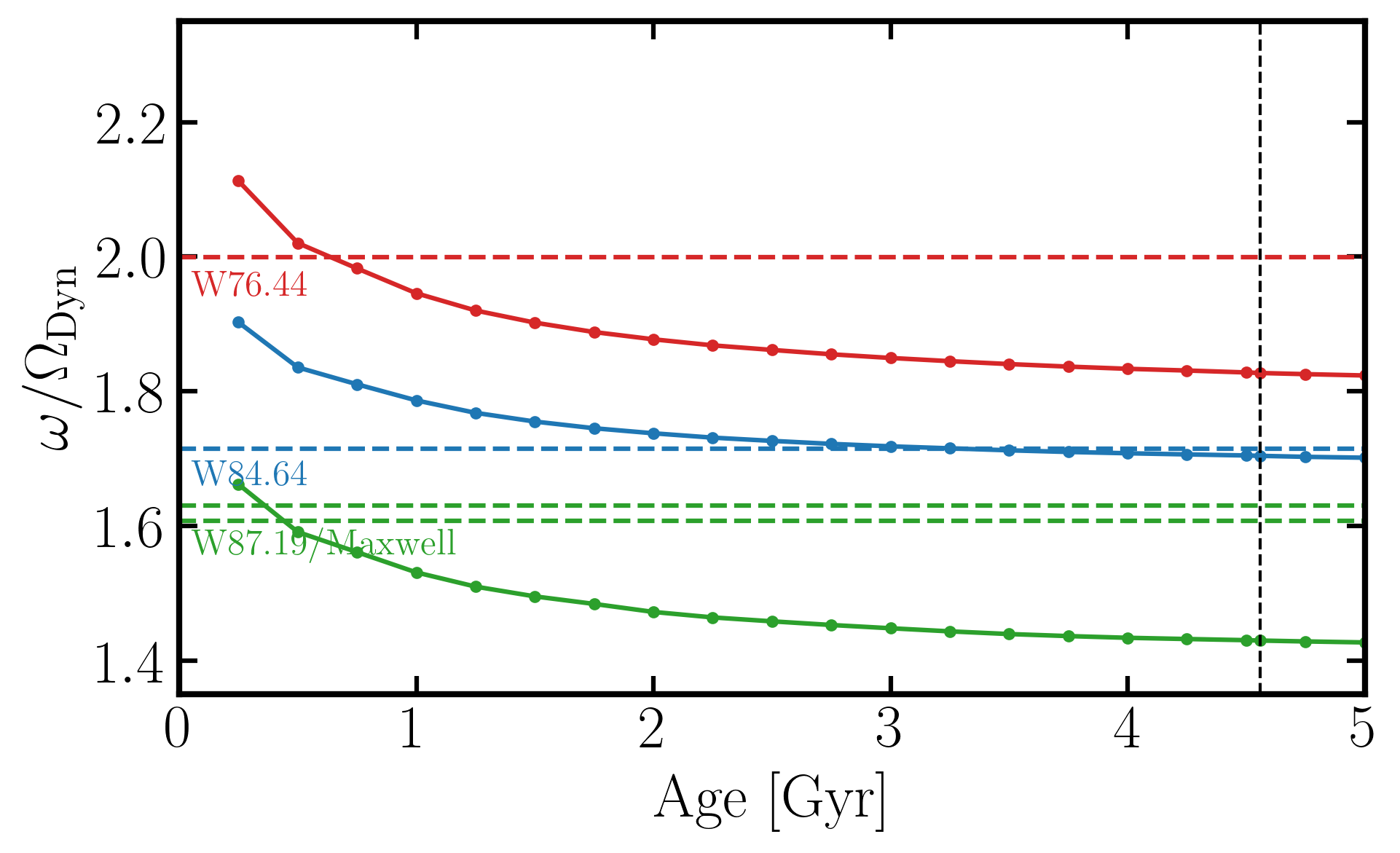}
    \caption{
    The top panel shows the surface gravitational potential perturbations for
    identified oscillation mode eigenfunctions for evolutionary model shown in
    Fig.~\ref{fig:sat_rotnocore} evaluated at $t=4.56\;\mathrm{Gyr}$,
    where the normalization is arbitrary but the relative amplitudes are correct
    under energy equipartition.
    The $f$ mode is denoted by the blue dot, and its inferred mode frequency
    from ring seismology is shown as the vertical blue dashed line.
    Similarly, the $n=1$ and $n=2$ $g$-modes are denoted by the red and green
    dots respectively, and are thought to correspond to the W76.44 ring
    oscillation pattern (vertical red dashed line) and the W87.19/Maxwell
    doublet (vertical green dashed lines).
    Transluscent black dots denote modes that are not spectrally resolved, but are illustrated here to highlight regions with rotationally-modified oscillations.
    The middle two rows of panels show color plots of the 2D gravitational
    potential perturbations ($\delta \Phi$) of the three identified modes,
    color-coded and labeled, and their most similar nearby modes (dotted boxes).
    Note that all of these nearby modes show distinct rosette mode-like
    features, illustrating the importance of consistent treatment of rotation in
    Saturn seismology.
    Finally, the bottom row shows the temporal evolution of the three identified
    mode frequencies; the vertical black dashed line denotes $t =
    4.56\;\mathrm{Gyr}$ as shown in the upper panels.
    }\label{fig:sat_rotnocore_surf}
\end{figure}

Next, we show the surface gravitational potential perturbations as a function of
the inertial-frame mode frequency for resolved modes in the searched frequency
range in Fig.~\ref{fig:sat_rotnocore_surf}.
We see that while a mode very near W84.64 appears (the $f$ mode of the planet),
no corresponding mode is found near the W76.44 frequency: the $n=1$ $g$ mode
is seen but at a lower frequency.
This suggests that the \brunt{} frequency in the deep interior of the planet is
insufficiently large to source an oscillation frequency compatible with the
W76.44 feature.
At the same time, the $n=2$ $g$-mode is seen at much lower frequencies than is needed to be relevant for the observed W87.19/Maxwell ring frequencies.

\subsection{With Helium Gradient}\label{ss:models1}

\begin{figure*}
    \centering
    \includegraphics[width=\textwidth]{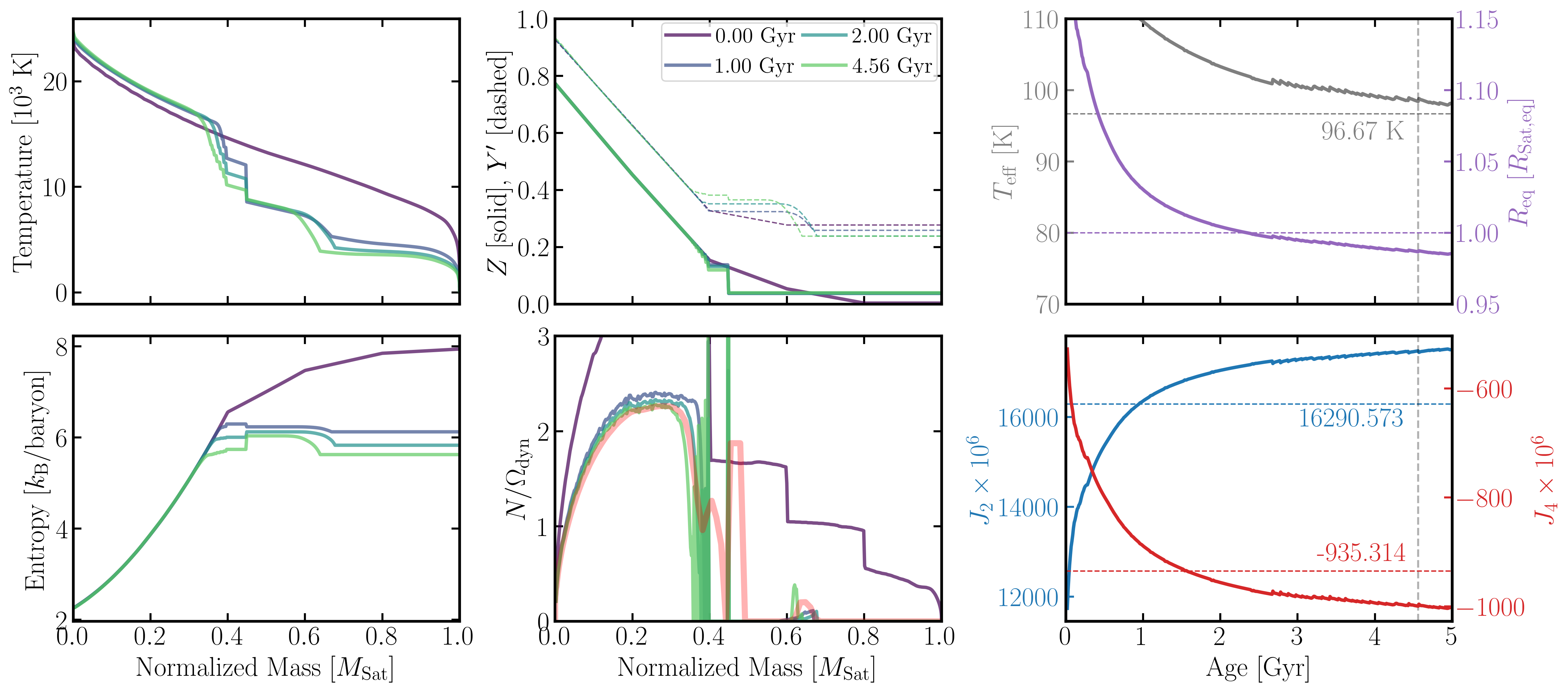}
    \caption{
    Same as Fig.~\ref{fig:sat_rotnocore} but for a model with a deep linear
    gradient in $Y' \equiv Y / (1 - Z)$ \citep[similar
    to][]{mankovich2021diffuse} and similar $Z$ and entropy profiles.
    An elevated \brunt{} frequency compared to the model shown in
    Fig.~\ref{fig:sat_rotnocore} can be seen due to the additional composition
    stratification, resulting in better agreement with the ring seismology
    constraints (see Fig.~\ref{fig:sat_ygrad11_surf}).
    Similar to Fig.~\ref{fig:sat_rotnocore}, the $J_2$ and $J_4$ values are
    elevated compared to their observed values.
    The final spin period is $0.995 P_{\rm Sat}$.
    }\label{fig:sat_ygrad11}
\end{figure*}

Next, in Fig.~\ref{fig:sat_ygrad11} we show a second model with a slightly
shallower initial heavy element gradient, but with a linear gradient in $Y'$ extending
over the same range as that of $Z$.
Our objective is to reproduce a profile like that of
\citetalias{mankovich2021diffuse} at the present day;
since our helium rain prescription does not give rise to helium differentiation
in stably stratified regions\footnote{
This may be physically sensible, as the differentiated fluid cannot efficiently
be transported throughout the planet.
}, a present-day $Y'$ gradient in the deep interior of the planet requires one
be there initially.
As in \citetalias{mankovich2021diffuse}, spatial coincidence of the composition
gradients is for convenience rather than being grounded in any physical
motivation.
The entropy gradient is adjusted to compensate for a comparable temperature
profile to that shown in Fig.~\ref{fig:sat_rotnocore}\footnote{
Note that our choice of $R_\rho = 1$ concerned purely the convective stability
condition and not the structural profile; in other work, we suggest the notation
$R_0$ for the structural density ratio in order to distinguish it from the ad
hoc convective prescription $R_\rho$ \citep{sur2025_rrho}.
Our profiles typically satisfy $R_0 \approx 0$, i.e., they are marginally
Schwarzschild stable
}.
The resulting evolution is shown in Fig.~\ref{fig:sat_ygrad11}, which contains a
total of $20 M_\oplus$ of heavy elements.
Comparable agreement with observational constraints is seen, where the radius is
too small and the gravity moments are too large.
The absolute values of $J_2$ and $J_4$ are too large, but as seen in
Fig.~\ref{fig:sat_rotnocore}, the magnitude of this discrepancy is sufficiently
small that it may be rectifiable by the addition of a compact core.
Note that \citetalias{mankovich2021diffuse} are able to obtain good agreement
with $J_2$ and $J_4$ in their original work.
In general, our models are hotter (less centrally dense) and have a higher
surface helium abundance, resulting in a larger $J_2$, despite nearly identical
radii and rotational periods.

\begin{figure}
    \centering
    \includegraphics[width=0.9\columnwidth]{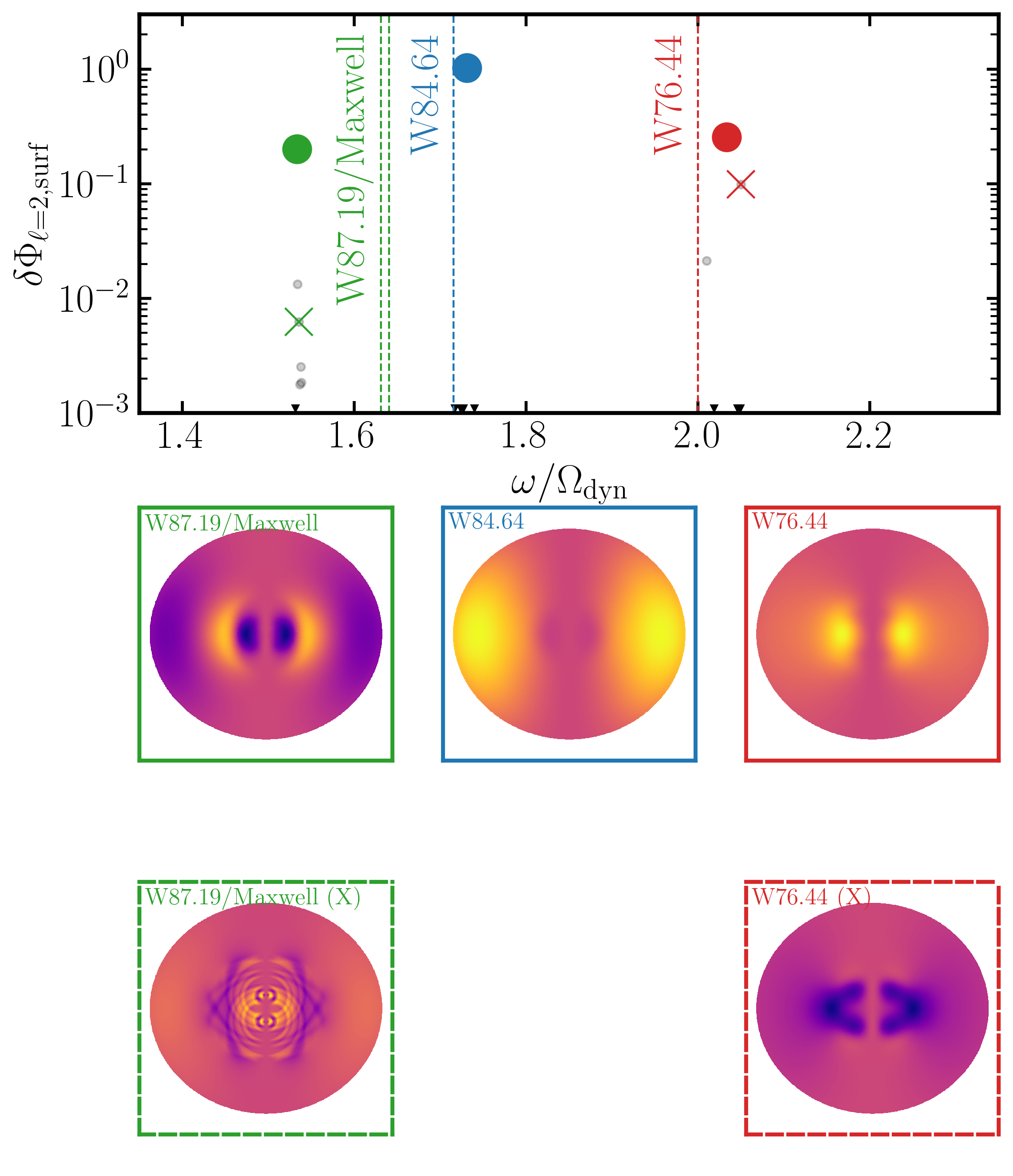}
    \includegraphics[width=0.9\columnwidth]{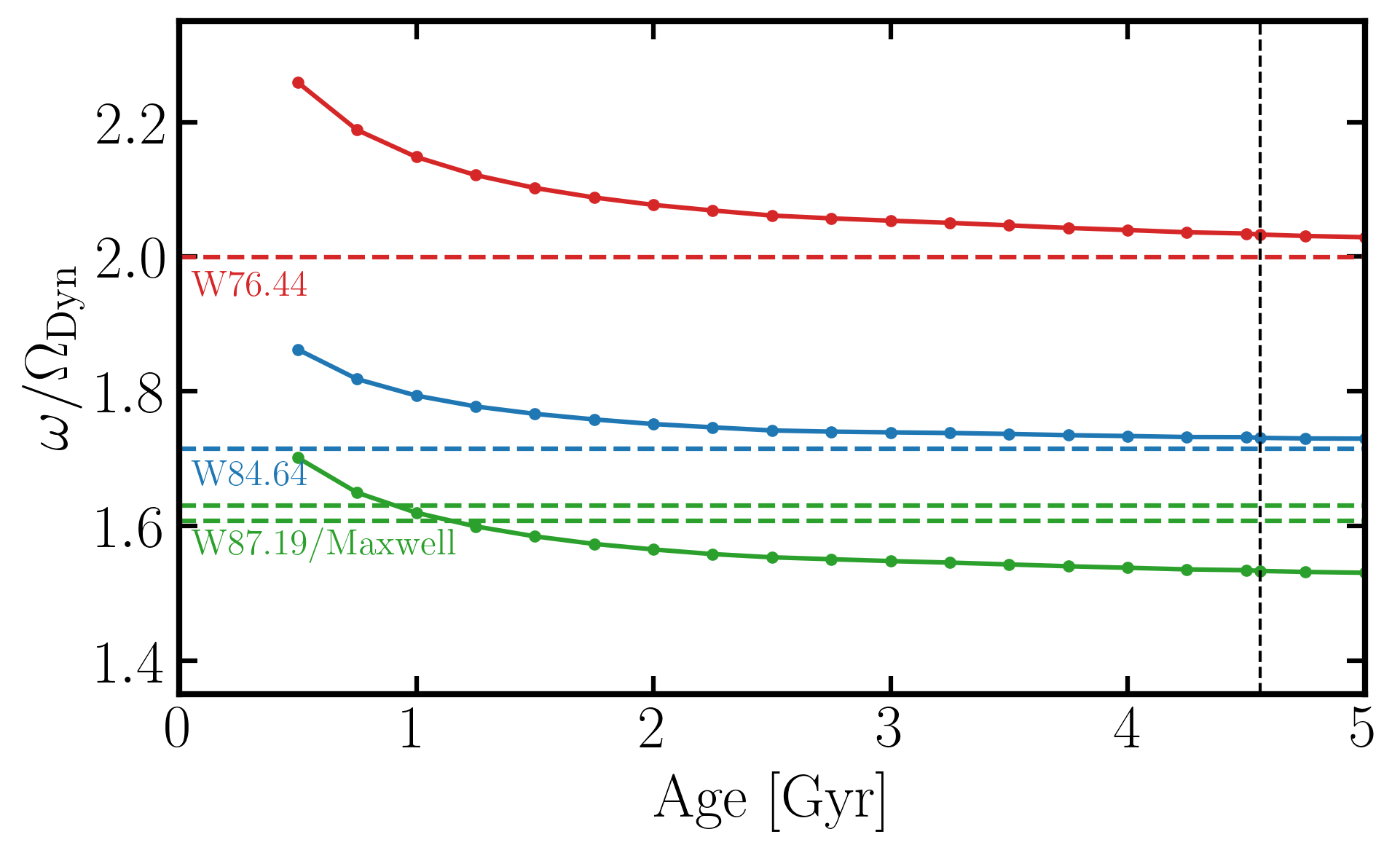}
    \caption{
    Same as Fig.~\ref{fig:sat_rotnocore_surf}, but for the model shown in
    Fig.~\ref{fig:sat_ygrad11} containing a $Y'$ gradient.
    The W84.64 and W76.44 frequencies are well-matched,
    while the W87.19/Maxwell doublet is not so well-matched.
    }\label{fig:sat_ygrad11_surf}
\end{figure}

On the other hand, this model is able to reproduce some of the
seismology, as can be seen in Fig.~\ref{fig:sat_ygrad11_surf}.
By comparison of the \brunt{} profiles obtained in Fig.~\ref{fig:sat_rotnocore}
and Fig.~\ref{fig:sat_ygrad11}, we see that the larger values of $N$ are
sufficient to attain the W76.44 mode frequency.
On the other hand, the outer extent of the stably stratified zone is $R_{\rm
stab} \approx 0.5R_{\rm Sat}$, modestly smaller than the $R_{\rm stab} \sim
0.6R_{\rm Sat}$ found by \citetalias{mankovich2021diffuse} (and also
corroborated by \citealp{mankovich2023_satrot2}).
This difference is primarily mandated by the efficiency of fuzzy core erosion in
our evolutionary models: As the surface cools, the convective exterior of the
planet grows (see \citealp{knierim2024_erosion} for an excellent theoretical
model describing the process).
Qualitatively then, the fact that Saturn must cool for $4.56\;\mathrm{Gyr}$
limits the maximum extent of the stably stratified interior region
(see Section~\ref{ss:core_erosion} for qualitative discussion about the
tightness of the limit derived in this work).
We recall that the W76.44 mode frequency is attributed to an avoided crossing
between the surface $\ell=2$ $f$ mode and an interior $n=1$, $\ell=2$ $g$ mode
\citep{mankovich2021diffuse}.
From our results, it appears that a reduced radial extent of the cavity still
produces a sufficiently strong avoided crossing to manifest a large surface
perturbation, comparable to that found by \citetalias{mankovich2021diffuse}.
However, the reduced cavity extent results in a lower frequency for the $n=2$
$g$-mode, such that it is unable to match the observed W87.19/Maxwell doublet
(note that the $n=2$ $g$-mode in the models of \citetalias{mankovich2021diffuse}
come closer to matching the W87.19/Maxwell doublet).
This cannot be attributed to uncertainties in the planet's cooling physics, as the
bottom panel of Fig.~\ref{fig:sat_ygrad11_surf} shows that there is no planetary age where all three frequencies are well reproduced.
We defer further interpretation and discussion of this discrepancy to
Section~\ref{ss:comp}.

We briefly comment that Fig.~\ref{fig:sat_ygrad11_surf} shows an additional mode
near the W76.44 that corresponds to a rosette mode \citep{takata2013_rosette,
dewberry2021_satrings} that undergoes an avoided crossing with the identified
$g$ mode.
The non-detection of a doublet at the W76.44 frequency in ring wave observations
suggests that either the rosette mode is less strongly forced / more strongly damped, or that
this particular model is disfavored.

Given the strong sensitivity of rosette modes to small features in the planet's
structure, it is difficult to draw concrete conclusions on whether small
modifications to the planet's structure will remove this doublet.
Empirically, we find that rosette modes near the W76.44 frequency appear in many
of our models, though their relative gravitational perturbations can easily be
reduced with small modifications to the planet's structure.

\subsection{Uniform Helium Concentration}\label{ss:models2}

\begin{figure*}
    \centering
    \includegraphics[width=\textwidth]{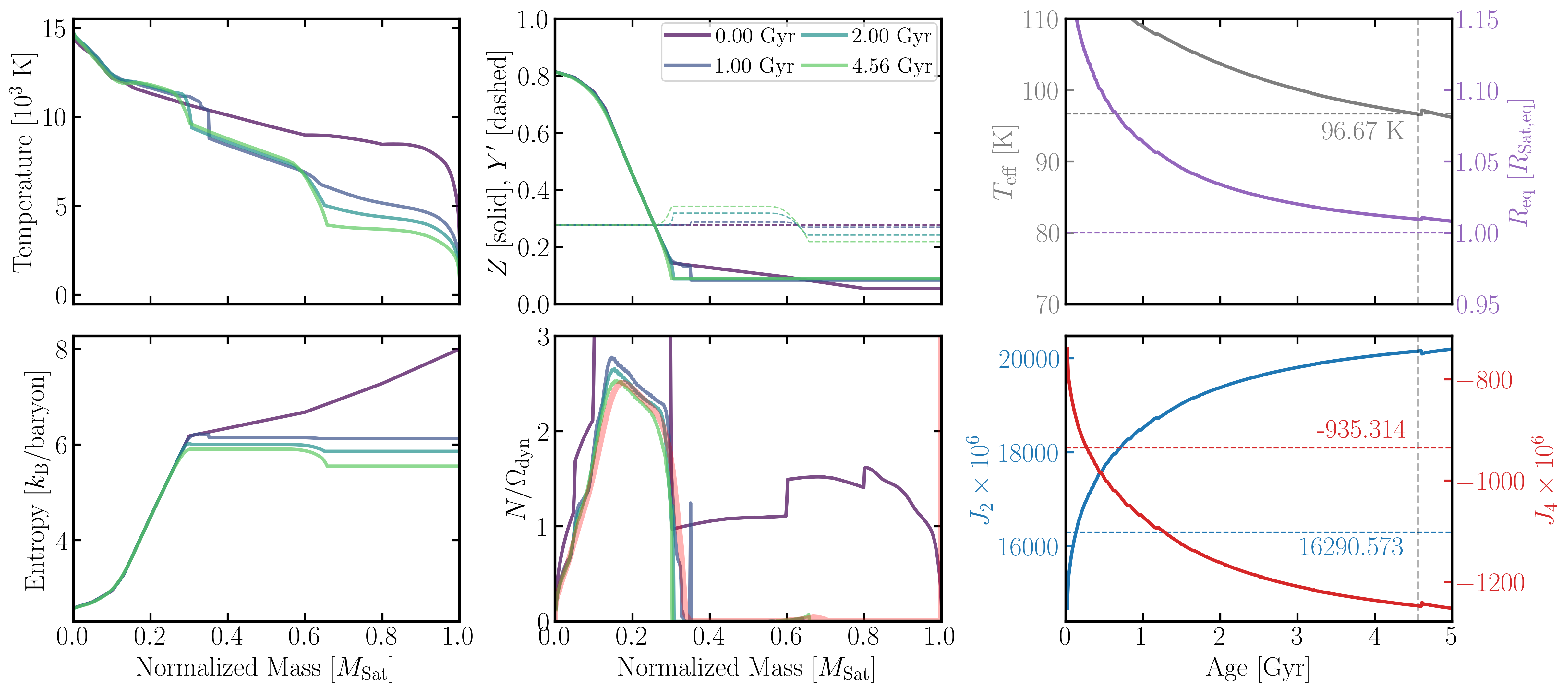}
    \caption{
    Similar to Fig.~\ref{fig:sat_rotnocore} but for a model with
    initially uniform $Y' \equiv Y / (1 - Z)$ and a steeper $Z$ profile.
    The final spin period is $0.994 P_{\rm Sat}$.
    Comparable agreement with observational constraints is seen in the right-most panels.
    }\label{fig:deepZ_hot}
\end{figure*}

While the model shown in Fig.~\ref{fig:sat_ygrad11} yields a mostly satisfactory
fit by evolutionary standards, it has one point of discomfort: why should there
be a primordial gradient in the helium composition $Y'$?
Naively, the initial
 $Y'$ should be set by the protosolar composition.
With this in mind, we study models with uniform initial $Y'$
profiles.
As seen from the results of Section~\ref{ss:models1}, a value of the
\brunt{} frequency $N \sim 2\Omega_{\rm dyn}$
 in the deep interior is required to admit a $g$ mode with a sufficiently high
frequency to match the W76.44 mode.
Since there is a smaller $Y$ gradient, and both $Y$ and $Z$ gradients contribute
to the value of $N$ \citep[cf.][]{tejada2024_eos}, a steeper $Z$ gradient is
required.
As such, we consider initial $Z$ profiles moderately steeper than those
considered in Section~\ref{ss:models1}.
With some small manual adjustments of the initial $Z$ and entropy profiles, we
arrive at the model shown in Fig.~\ref{fig:deepZ_hot}, which contains a total of
$22M_\oplus$ of heavy elements.
Note that the \brunt{} profile is somewhat larger in magnitude and less extended.

\begin{figure}
    \centering
    \includegraphics[width=0.9\columnwidth]{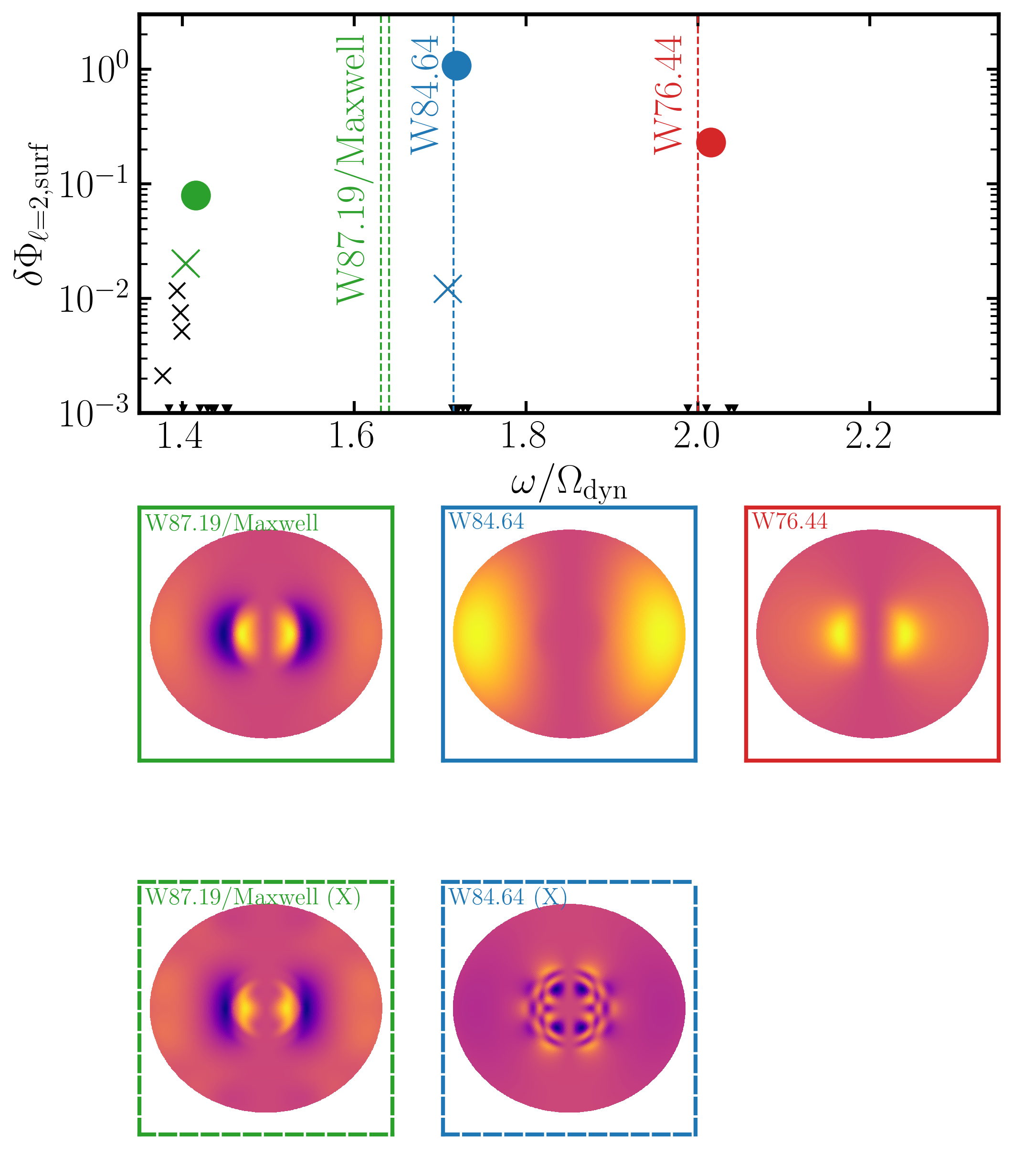}
    \includegraphics[width=0.9\columnwidth]{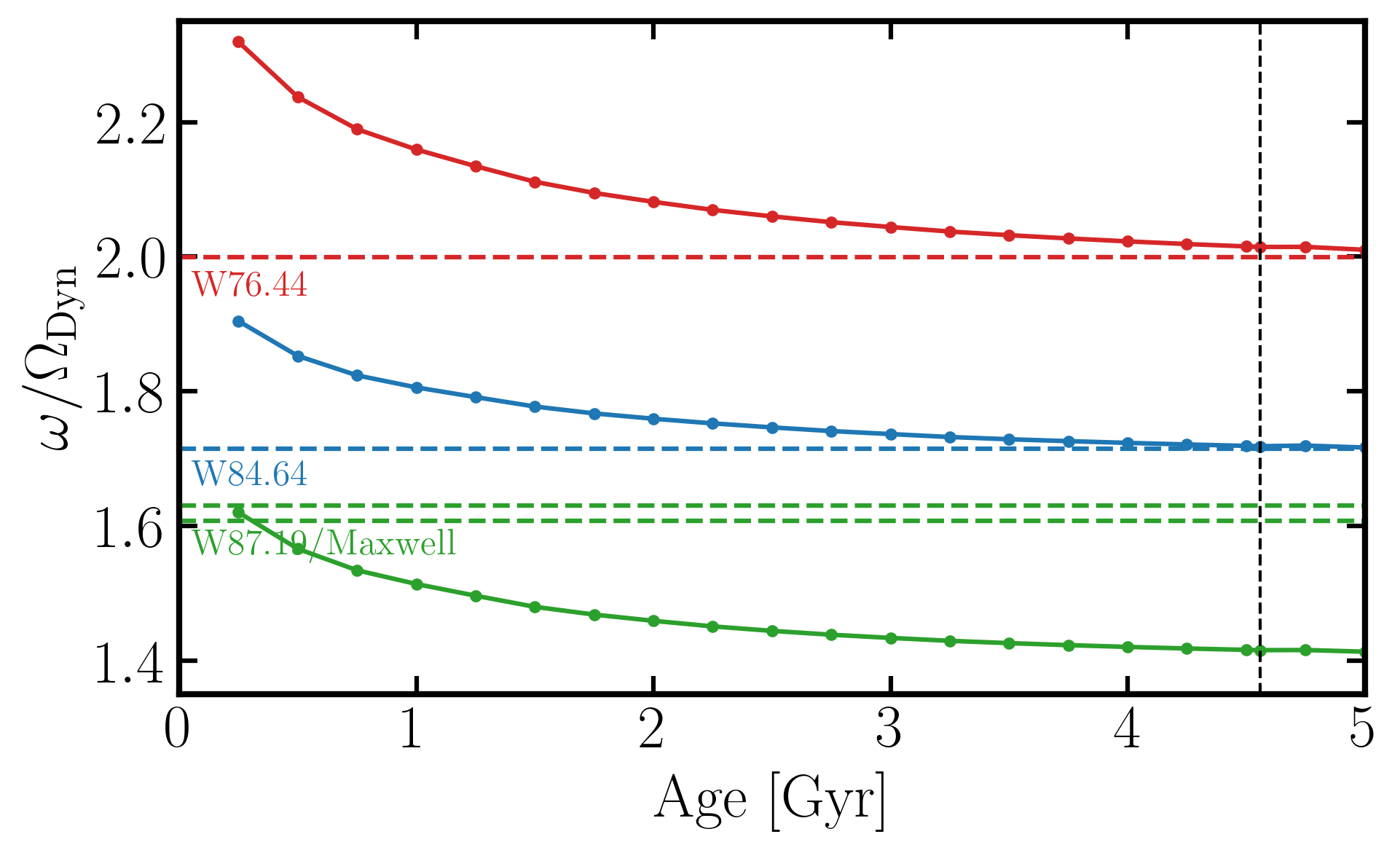}
    \caption{
    Seismology for a model corresponding to Fig.~\ref{fig:deepZ_hot}. Excellent
    agreement is observed for W84.64 and W76.44, though again the $n = 2$
    $g$-mode does not match the W87.19/Maxwell frequency.
    }\label{fig:deepZ_seis}
\end{figure}

The corresponding seismology can be seen in Fig.~\ref{fig:deepZ_seis}.
The agreement
of the $f$-mode and $n=1$ $g$-mode
is excellent and is marginally better than the model with a $Y'$
gradient studied above (Fig.~\ref{fig:sat_ygrad11_surf}).
Note also that while the radius of the inner stable region is only $R_{\rm stab}
\approx 0.43R_{\rm Sat}$, the surface gravitational perturbation of the $n=1$
$g$-mode is not significantly affected.
However, the small cavity size results in a significantly lower frequency for
the $n=2$ $g$-mode.

\begin{figure*}
    \centering
    \includegraphics[width=\textwidth]{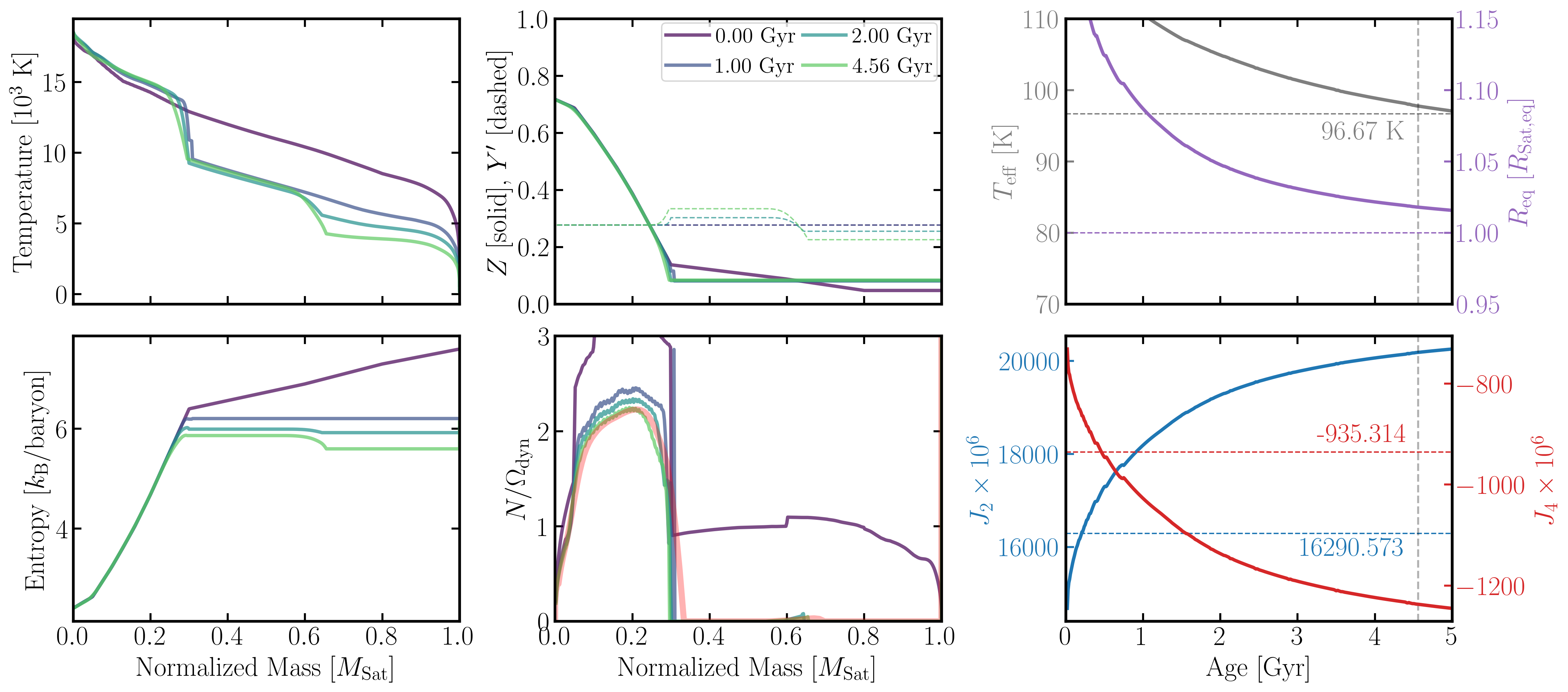}
    \caption{
    Similar to Fig.~\ref{fig:deepZ_hot} but for different profiles adapted for
    a postperovskite heavy element equation of state \citep{zhang2022_ppv}.
    While the planetary structure is quite different, the resulting agreement
    with observables is quite similar.
    The final rotation period is $0.989 P_{\rm Sat}$.
    }\label{fig:ppv}
\end{figure*}

\begin{figure}
    \centering
    \includegraphics[width=0.9\columnwidth]{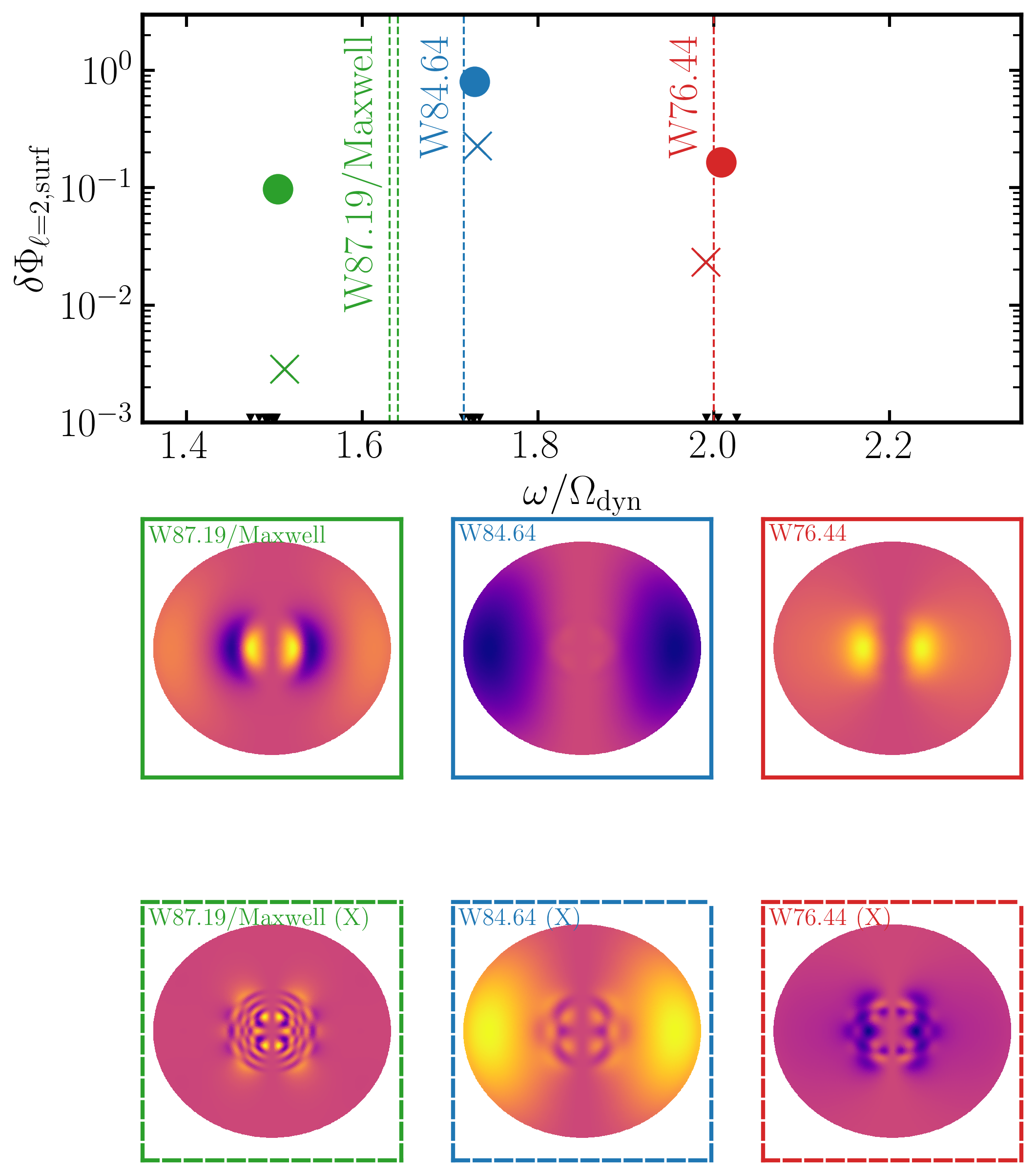}
    \includegraphics[width=0.9\columnwidth]{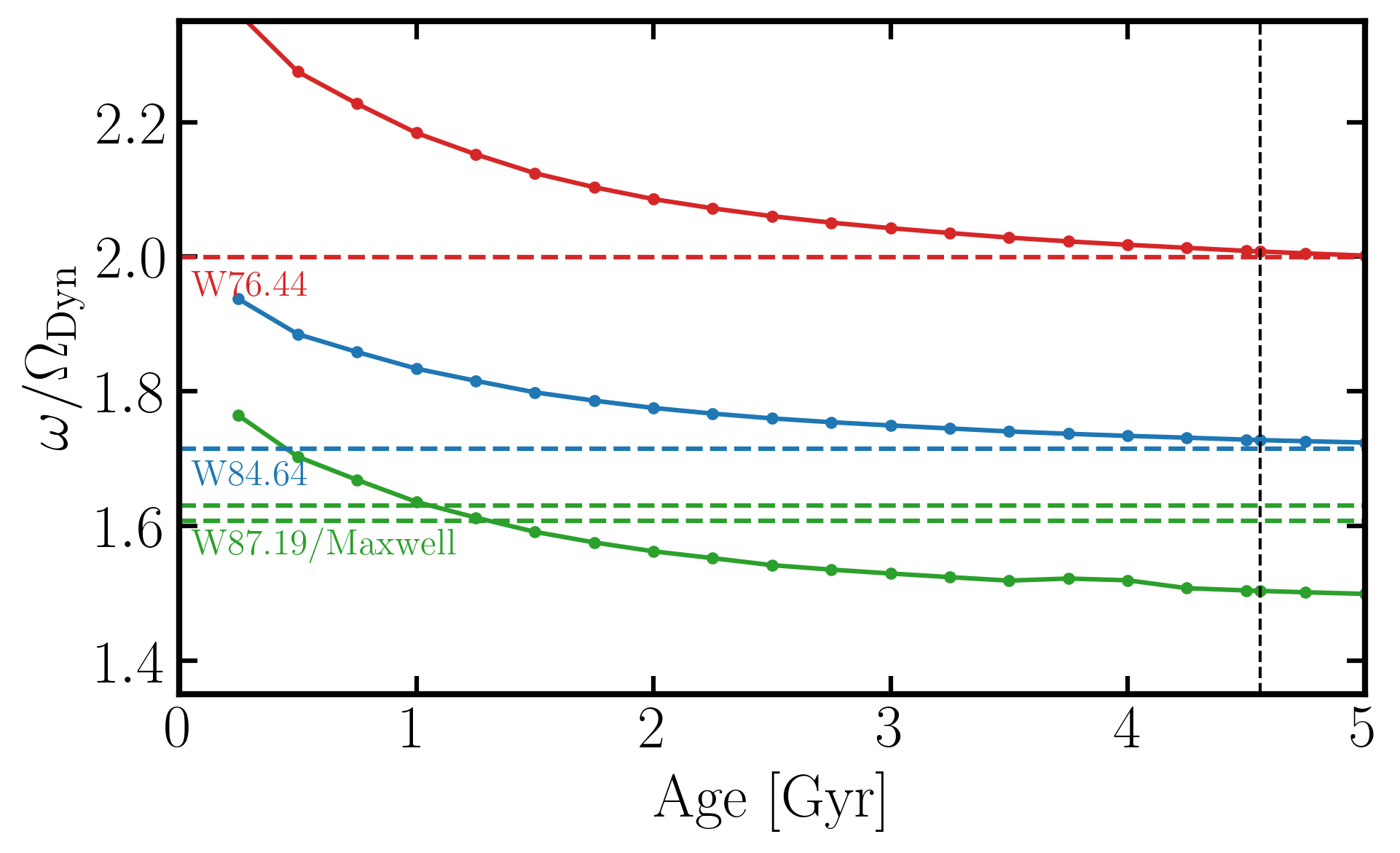}
    \caption{
    Seismology for model corresponding to Fig.~\ref{fig:ppv}.
    Modestly improved agreement of the $n=2$ $g$-mode with the W87.19/Maxwell is observed compared to Fig.~\ref{fig:deepZ_seis}, suggesting that composition may play a role in the detailed frequencies of higher-order $g$-modes.
    }\label{fig:seis_ppv}
\end{figure}

To explore the effect of uncertainties in the heavy element equation of state on
the inferred structures and observables, we also consider models where the
heavy elements are treated using a postperovskite equation of state
\citep[MgSiO$_3$;][]{keane1954_ppv, stacey2004_ppv, zhang2022_ppv}.
An example model retaining satisfactory agreement with the seismology is shown
in Fig.~\ref{fig:ppv}, which contains a total of $19M_\oplus$ of heavy elements.
This model has inner stable region extent $R_{\rm stab} \approx 0.44 R_{\rm
Sat}$.
Note that the overall heavy element mass fraction ($Z$) in the deep interior is
lower.
This is because postperovskite is denser at similar temperatures than water, so
when using similar temperature profiles, stronger stratification and larger
\brunt{} frequencies are produced.
As such, when the target \brunt{} profile is held fixed, a shallower $Z$
gradient arises for postperovskite than for water.
The agreement with the known seismological frequencies remains quite good, as
can be seen in Fig.~\ref{fig:seis_ppv}.
Note that the $n=2$ $g$-mode is now found at a higher frequency than in the AQUA
model shown in Fig.~\ref{fig:deepZ_seis}, though mixing with a nearby
rosette mode significantly suppresses its $\delta \Phi_{\ell=2, \mathrm{surf}}$.
Given that both models have similar $R_{\rm stab}$, the differing frequencies of
the $n=2$ $g$-mode may be due to the different shapes of the \brunt{} frequency
profiles.

\section{Discussion}\label{s:disc}

We have shown that good simultaneous agreement with most of the listed
properties in Table~\ref{tab:params} can be achieved both with a primordial $Y'$
gradient (Section~\ref{ss:models1}) and without (Section~\ref{ss:models2}), with
the notable caveat of failing to reproduce the
gravitational harmonics $J_2$ and $J_4$.
By comparison to our previous modeling efforts, we suggest that these
discrepancies are due to the absence of a rocky core in the models considered
here.
We anticipate that similar structures to those presented here can be found that
both have rocky cores and fit all of Saturn's observables, though future
numerical work is necessary to confirm this.
Below, we briefly comment on a few minor caveats of the evolutionary treatment
in this work, then we discuss our results in the context of previous work in the
literature.

As a first caveat to our evolutionary approach, we have not considered
semi-convective energy and composition transport in this work.
\citet{sur2025_rrho} found that a low value of the $R_\rho$ convection
parameter, corresponding to efficient semiconvective transport, results in
greater convective mixing and a smaller fuzzy core.
However, the $R_\rho$ prescription does not truly model semiconvective
processes, so its conclusions should be taken with some degree of caution;
detailed numerical work will be required to explore the structural effect
of more physical treatments of semiconvection.
Note that, if the stably stratified region is semi-convective, it may give rise
to magnetic dynamos \citep{pruzina2025_semiconvdynamo} that help reproduce the
observed axisymmetry of Saturn's magnetic field \citep{stevenson1982_symmdynamo,
stanley2010_dynamo, yan2021_dynamo}.

It has been suggested that a helium compositional gradient due to immiscibility
may give rise to stable stratification \citep[e.g.,][]{guillot2004_jup,
markham2024_rain}.
It is then natural to wonder whether this stratification results in substantial
changes to the seismological results.
In our models, we find that even when helium demixing drives the formation of a
stably stratified region, the effect of this region on the $\ell=2$ oscillation
spectrum is small.
We provide an analytic argument justifying this lack of an effect in
Appendix~\ref{s:rain_rrho}.

Finally, we contextualize the temporal uncertainty underlying the models we
display at $t=4.56\;\mathrm{Gyr}$.
Recent working using \texttt{APPLE} has found that the surface boundary
condition can affect model radii and surface temperatures at the $\sim 3$--$7\%$
level \citep{sur2026_exoplanet}, while different prescriptions for
semiconvective energy transport (via the $R_\rho$ parameter) can affect the
radius evolution by tens of percent \citep{sur2025_rrho}.
Taken together, the implication is that the time it takes for a planet to reach
a given structure in evolutionary models could be uncertain by anywhere from a
few to a few tens of percent, depending on the modeling assumptions.
Nevertheless, as can be seen in the bottom panels of
Figs.~\ref{fig:sat_ygrad11_surf},~\ref{fig:deepZ_seis}, and~\ref{fig:seis_ppv},
even uncertainties in the evolutionary age at this level does not significant
affect the mode frequencies and the conclusions of this paper.

\subsection{Comparison to Previous Work}\label{ss:comp}

We begin with an in-depth comparison between previous static structural modeling
of Saturn incorporating ring seismology (\citetalias{mankovich2021diffuse},
\citealp{dewberry2021_satrings, dewberry2022_saturnwinds,
mankovich2023_satrot2}) and our current work.
Both their models and ours yield comparable estimates for Saturn's rotation
rate, radius, and two of the ring seismology modes.
However, their models are additionally able to arrive at satisfactory (yet not
quite perfect) fits to the gravity moments $J_2$ and $J_4$ fit the $n=2$
$g$-mode to the W87.19/Maxwell doublet.
In order to do so, they impose a helium gradient ($Y'$) throughout the stably
stratified region and require a rather extended stably stratified region
($R_{\rm stab} \sim 0.6 R_{\rm Sat}$).
By contrast, our evolutionary models find a less extended stably stratified
region ($R_{\rm stab} \approx 0.4$--$0.5 R_{\rm Sat}$) due to convective mixing
over time, and as a result are unable to successfully match the $J_2$, $J_4$, or
the W87.19/Maxwell doublet.
We do find that these constraints are all somewhat better fit in the presence of
a $Y'$ gradient (compare the models in Section~\ref{ss:models1}
and~\ref{ss:models2}).
However, in our models, such a gradient must be present at formation, as our
helium rain model does not produce composition gradients in stably stratified
regions (which we propose is physically accurate); such an initial condition
seems ill-motivated physically.

First, we remark that the difference in stable region extent is driven by
evolutionary constraints.
We qualitatively outline the constraining effects below: unfortunately, rigorous
quantitative bounds cannot be estimated analytically and can only be determined
via detailed numerical exploration of the nature here.
In our models, the combined constraints of Saturn's present-day radius and that
enough heavy elements be present in the deep interior to source a large \brunt{}
frequency place a lower bound on Saturn's initial entropy profile Saturn---if
the planet is too cool, it must either have a small present-day radius or low
concentrations of heavy elements.
On the other hand, an imposed minimum entropy translates into a minimum
luminosity; since convective mixing is directly related to surface luminosity
\citep{knierim2024_erosion}, these constraints translate to a lower limit on the
total amount of convective mixing, and an upper limit on $R_{\rm stab}$.
This constraint can be relaxed somewhat by decreasing the total heavy element content,
the planet's initial temperature,
and introducing a primordial $Y'$ gradient to make up for the decrease to the
\brunt{} frequency.
However, again, the physical motivation for such an initial condition is tenuous
at best.

Second, we remark that the excessive $J_2$ and $J_4$ values in our models are
largely due to the smaller $R_{\rm stab}$ values in our models.
We find that in evolutionary models where the present-day $R_{\rm stab}$ is
larger, we successfully reproduce the observed $J_2$ and $J_4$ values even in
absence of a rocky core.
However, due to the constraints discussed above, such evolutionary models fail
to fit either the radius or seismology results.
Motivated by our previously published Saturn models that required a compact
rocky core \citep[as shown in
Fig.~\ref{fig:sat_rotnocore},][]{sur2025_fuzzycores}, we suggest that the
addition of a rocky core to evolutionary models will reduce $J_2$ and $J_4$ (by
$\sim 25\%$) while preserving simultaneous fits of all observed properties of
Saturn.
Note that \citetalias{mankovich2021diffuse} also considered models with a
``core'' corresponding to a region of constant $Z_{\rm in}$ interior to some
inner radius $r_{\rm in}$.
However, in order to ensure continuity of their $Z$ profiles, their $Z_{\rm in}
\sim 0.65 < 1$\footnote{
Specifically, they find that higher $Z_{\rm in}$ require larger ice fractions
while lower $Z_{\rm in}$ require larger rock fractions. However, they adopt the
ANEOS serpentine and water ice equations of state \citep{thompson1990aneos},
which differ from the AQUA and ppv ones considered here.
}.
By contrast, the rocky cores investigated in our previous evolutionary models of
Saturn \citep{sur2025_fuzzycores} have a discontinuous jump in the $Z$ profile
such that $Z_{\rm in} = 1$.
Since a steeper jump in $Z_{\rm in}$ results in a greater degree of central mass
concentration and lower $J$ values, a $Z_{\rm in} = 1$ core will have a greater
effect on $J$ values than a $Z_{\rm in} < 1$ core\footnote{
We attempted to model such a compact core with a smoothed $Z$ profile,
but the resulting \brunt{} profiles were far too large in the transition region
between the compact rocky core and the extended fuzzy outer region, affecting
the ring seismology.
With a sharp boundary between these two regions, this problem may be mitigated,
since the $g$-modes of the fuzzy core remain trapped exterior to the rocky core;
future work will provide concrete answers to this possibility.
}.
Such a sharp boundary may naturally arise during the formation of Saturn
\citep{bodenheimer2025_saturn}.
Given the ability of the \citetalias{mankovich2021diffuse} models with $r_{\rm
in} > 0$ to still fit the data, a guarded degree of optimism that $Z = 1$ rocky
cores can yield agreement with the seismological data seems warranted.
Finally, note that the reduced values for $J_2$ and $J_4$ will also reduce the
need for the large $Y'$ gradient inferred by \citetalias{mankovich2021diffuse},
which is primarily motivated by the combined need for a large \brunt{} frequency
and a large degree of central mass concentration.

Third, our models with and without primordial $Y'$ gradients yield comparable
fits to the seismological results (see Figs.~\ref{fig:sat_ygrad11_surf}
and~\ref{fig:deepZ_seis}):
both are able to fit the W84.64 and W76.44 frequencies quite well, but are
unable to fit the W87.19/Maxwell doublet.
However, the agreement with the W87.19/Maxwell doublet may be dependent on the
heavy element composition and equation of state:
comparison of the models using AQUA and post-perovskite show that, when equally
good agreement with the W84.64 and W76.44 modes is required, the $n=2$ $g$-mode is
located at a higher freequency for the post-perovskite equation of state (though
the dependence is somewhat weak, in accordance with the findings of
\citealp{mankovich2021diffuse}; see their Extended Data Fig.~4).
This comparison may be moot, however, if the W87.19/Maxwell doublet is due to
other oscillations than the the $n=2$ $g$-mode \citep[e.g., some combination of
$\ell\sim2,4,6$ g-modes or rosette modes; ][]{dewberry2021_satrings}.
We conclude that there appears to be be some inherent degeneracy between the
heavy element composition and its concentration/distribution in Saturn, though
there may be prospects for reducing the degeneracy in future work that
incorporates the complete spectrum of detected $f$-modes, as well as the
additional $|m|=1,3$ ring waves that have yet to be unambiguously linked to
specific modes in Saturn (\citealp{hedman2013kronoseismology,
hedman2014krono2}, but see \citealp{friedson2023saturn}).

Lastly, we
comment on the adiabaticity of the stratification considered in
these models.
The initial entropy profiles we explore are created without particular regard
for the adiabaticity of the stratification, though we verify \emph{a posteriori}
that our profiles are approximately adiabatic.
\citetalias{mankovich2021diffuse} find that the superadiabaticity of
the temperature profile does not significantly affect the \brunt{} frequency
profile, even if the core temperature changes by a factor of up to $\sim
3$.
At the same time, while near-adiabatic stratification (with density ratio $R_0 =
R_{\rm crit} \sim 10^{-2}$) seems somewhat appealing as the equilibrium
temperature profile under semiconvection \citep{wood2013_ddconv,
garaud2018_ddiff}, numerical simulations suggest that compositional transport
under semiconvection is $\sim 10\times$ more efficient than thermal transport
\citep{wood2013_ddconv, moll2016_dd}.
As such, the ability of semiconvection to shape the temperature profile in the
stably stratified region without destroying its compositional stratification
seems to require additional explanation, so a near-adiabatic temperature
profile may neither be necessary nor hold any particular significance.

In summary, our evolutionary models differ from previous static structural
inferences of Saturn's interior, primarily due to its smaller extent of central
stratification.
By doing so, our models fail to predict the $J_2$, $J_4$, and yield no $g$-modes
near the W87.19/Maxwell doublet.
We suggest that some of these shortcomings may be mitigated by the addition of a
compact rocky core.

\subsection{Convective Mixing Processes}\label{ss:core_erosion}

The accuracy of the combined evolutionary and seismological results in this
work hinges critically on the extent of the stably stratified core that is able to survive to
the present time.
However, there have been many recent, highly detailed works on the effect of rotation on the survivability
of extended stably stratified regions in giant planet interiors \citep{fuentes2023_erosion, hindman2023_rotationconvgrowth}.
Here, we discuss the relation between these advances and the overall process of convective mixing\footnote{
Note that convective mixing is often referred to in other works as ``core erosion.''
We take care to avoid the term erosion, which we think subliminally suggests
that the processes driving the shrinkage of stably stratified cores is
exclusively due to fluid motions that dig into the core.
}, and we conjecture on their possible bearing on Saturn's interior structure inference.
%

In this current work \citep[and many others preceding it,][]{
knierim2024_erosion, tejadaarevalo2025_jupiterfuzzy, sur2025_fuzzycores},
the gradual mixing of the initially stably-stratified interior can be
understood by a direct analysis of the energy transport in the planet.
When the outer regions of the planet cool due to radiation from the surface,
they eventually become sufficiently cooler than the immediately interior fluid
that the fluid at the boundary becomes unstable to convection.
This instability is powered by the excess thermal energy of the interior of the planet relative to the cooling surface.
Then, as the planet continues to cool, the entropy of the isentropic convective envelope decreases, and the radiative-convective boundary moves inwards (to lower entropies).
Since this explanation only depends on the radial thermodynamic profiles
of the planet, they manifest even in one-dimensional
planetary evolution studies.
However, convection itself is a multi-dimensional fluid instability,
and crucially the flow structures change in the presence of rotation,
resulting in modified flow structures and energy transport
\citep{augustson2019_rmlt, stevenson1979_cia}.
Recent 3D hydrodynamical simulations have suggested that the effect
of this rotationally-modified transport is a marked decrease of convective
mixing efficiency in giant planet interiors \citep{fuentes2023_erosion,
hindman2023_rotationconvgrowth, zhang2025_fuentes_hydrosimentrainment}.
While neither the 1D simulations nor the 3D simulations are fully physically realistic,
the consequences of convective mixing are significant for modeling
the evolution of giant planet interiors.
Below, we assess the likely implications of the physical effects missing in
our work and conclude that our conclusions may be physically faithful despite
these limitations.

First, we point out the key feature of the mechanism studied in the current work: given an initial planetary profile (entropy and composition), the total energy emitted fully constrains the final extent of the interior stably stratified region:
While no simple analytic expression can be derived, the constraint of near-isentropy and near-homogeneity in the outer region of the planet is sufficient to enforce a one-to-one mapping between integrated luminosity and planet structure \citep{knierim2024_erosion}.
However, an important correction imposed by rotation is that the efficiency of
convective energy transport is reduced \citep{stevenson1979_cia,
augustson2019_rmlt}.
As such, for identical initial conditions, a rotating planet will equilibriate
to a slightly steeper entropy gradient compared to a non-rotating planet, and
its surface will be accordingly cooler than predicted in the current work.
However, this effect appears to be minor:
The temperature discrepancy can be estimated using rotational mixing length
theory (RMLT), under which the superadiabaticity in convective flows scales
like $\propto \mathrm{Ro}^{-4/5}$ \citep{agrawal2026_twistsintheflow}, where
the Rossby number $\mathrm{Ro} \sim 10^{-6}$ in giant planet interiors
\citep{guillot2004_jup}.
To estimate the effect that this has on the total emitted energy of the planet,
we report that the typical entropy difference across the convective region in
our models is $\sim 10^{-7}\;\mathrm{k_{\rm B}/baryon}$.
As such, under RMLT, the surface entropy may be reduced by $\sim 10^{24/5}
\times 10^{-7} \sim 0.01\;\mathrm{k_{\rm B}/baryon}$ in the presence of
rotation, equating to a change in the internal temperature of $\sim
2\;\mathrm{K}$.
This is equivalent to a moderate ($\lesssim 200\;\mathrm{Myr}$ depending on the
age) change in the inferred age of the planet.
More importantly, the effect on the total radiated energy is similarly modest,
leading to small differences to planetary interior structures with and
without rotational modifications to convection.
While this analysis is coarse, and a full implementation of rotational mixing
length theory in our updated planet evolution code \texttt{ORCHARD}
\citep{tejada2026_orchard} is in progress, it suggests tentatively that
the omission of rotationally-suppressed convective transport is not a dominant
source of uncertainty in the planetary structures presented here.
At the same time, further hydrodynamical studies with more realistic physical
parameters and prescriptions will be needed to bridge the gap between these two
pictures.

\section{Summary}\label{s:summary}

A significant challenge to the modern understanding of giant planet formation
is the discovery of extended stably stratified ``fuzzy cores.''
The strongest constraints on the extent of these fuzzy cores are obtained
by seismological inference of features in the rings of Saturn
\citep[][]{mankovich2021diffuse}.
The presence of these extended fuzzy cores poses a challenge for
planetary evolution models, which find that the extent of fuzzy cores
should generally decrease as the planet evolves, for a mixture of
energetic and hydrodynamic reasons.

In this work, we have used giant planet evolutionary models to understand the
balance between these two competing demands.
Our work shows that these two effects can be reconciled: a sufficiently
extended stably stratified core consistent with constraints from ring seismology
can survive to the present day.
Our models generally require steeper stratification than evolutionary models
in the literature today, but suggest less extended fuzzy cores compared to static model inferences from ring seismology \citep{mankovich2021diffuse}.
Our combined approach and results illustrate the importance of ring seismology towards constraining planetary interior evolution.
Furthermore, by comparing our models that exhibit good fits to the observed properties of Saturn, we conclude that the possible presence of a compact rocky core are the most promising avenues for
improved fits to Saturn's present-day structure.
We defer exploration of these effects to future work.

\section*{Acknowledgements}

We thank the referee Chris Mankovich for a thorough, thoughtful,
and exacting review that significantly improved the quality of this paper.
We thank Jim Fuller, Linhao Ma, and Jos\'e Rafael Fuentes for illuminating
discussions.
YS acknowledges support by the Lyman Spitzer Jr.\ Postdoctoral Fellowship at
Princeton University and by the Natural Sciences and Engineering Research
Council of Canada (NSERC) [funding reference CITA 490888-16]. Partial funding
for this research was provided by the Center for Matter at Atomic Pressures
(CMAP), a National Science Foundation (NSF) Physics Frontier Center, under Award
PHY-2020249. Any opinions, findings, conclusions or recommendations expressed in
this material are those of the
author(s) and do not necessarily reflect those of the National Science
Foundation.

\bibliography{refs}{}
\bibliographystyle{aasjournalv7}

\appendix

\section{Relevance of Helium Rain for Seismology?}\label{s:rain_rrho}

Helium rain introduces a helium gradient across the immiscibility region.
It has accordingly been suggested that this region may be stably stratified,
which may then play a role in the interpretation of Saturn's seismology
\citep[e.g.,][]{mankovich2021diffuse}.
In the course of our work, we found that the rain region generally does not produce
stably stratified regions that affect the crucial W76.44 frequency.
In this section, we provide an analytical argument for why this is the case.

In the immiscible region, the thermodynamic profile remains marginally convectively
unstable throughout.
This is because any stable regions cease to efficiently transport helium and energy,
resulting in significant accumulation of helium and heat beneath the stable region
that quickly overwhelms the stabilizing gradients.
Thus, in the absence of semiconvection, the rain region should remain fully convective.
This is largely borne out by the results shown here (e.g., Figs.~\ref{fig:sat_ygrad11}).
Note that a small \brunt{} frequency can be seen in the immiscible region.
This is because in the operator splitting adopted by \texttt{APPLE}, each timestep
consists of a thermodynamic adjustment followed by a hydrostatic one.
Thus, even though the rain region is fully convective after the thermodynamic adjustment
(verified by inspection), it can become marginally convective after the hydrostatic
adjustment.

To then understand the circumstances under which stable stratification due to
helium rain can arise, some semiconvective transport must be adopted.
To qualitatively understand the effect of semiconvective transport,
we adopt the $R_\rho$ prescription \citep{mankovich2016_rrho, sur2025_rrho},
which we briefly summarize below.
Note that our results will qualitatively apply to more physical models
of semiconvective transport as well (e.g., $R_\rho = 0$ corresponds to
highly efficient semiconvection).

In \texttt{APPLE}, the $R_\rho$ prescription is implemented as a modification to
the convective velocity \citep{mankovich2021diffuse, sur2024_apple,
sur2025_rrho}
\begin{align}
    v_{\rm conv}
        \propto
            \Bigg\{& -\rd{S}{r}
            + \Bigg(
                \p{\pd{S}{Y}}_{PT}(1 - R_\rho)
                + \p{\pd{S}{Y}}_{P\rho}R_\rho
            \Bigg)
            \rd{Y}{r}
                \Bigg\}_+^{1/2},\label{eq:vconv}
\end{align}
where the $+$ subscript denotes taking the maximum of the bracketed quantity and
zero.
In the limit $R_\rho = 1$, convective energy transport sets in only once the
planetary fluid becomes unstable under the Ledoux condition \citep{ledoux1947}.
On the other hand, in the limit $R_\rho = 0$, convective energy transport
becomes possible as soon as the planetary fluid is unstable under the
Schwarzschild condition (see \citealp{sur2025_rrho} for a more detailed
discussion), modeling efficient energy transport via semiconvective processes
such as double diffusive instabilities \citep[e.g.][]{leconte2012_doublediff,
leconte2013_ddiffusive, garaud2018_ddiff}.
We explicitly remark that the physical convective stability of the fluid, used
to calculate the \brunt{} frequency, is always evaluated using the Ledoux
condition regardless of the value of $R_\rho$ adopted.

A nonzero $N$ in the helium rain region appears when $R_\rho < 1$ is taken,
since the effective convective flux ($\propto v_{\rm conv}^3$) can be nonzero
even when the fluid is physically convectively stable.
In our numerical experiments, the stably stratified regions due to helium rain
were unable to support large $g$ mode frequencies, and we provide a qualitative
justification for this here.
Recall that standard expressions for the $g$ mode frequencies typically scale as
\begin{equation}
    \omega_g^2 \simeq \frac{N^2 l(l+1)}{r^2 k_r^2}\, ,
\end{equation}
where $r$ is the radius of the stably stratified cavity and $k_r$ is the mode
radial wavenumber. The largest frequencies then correspond to the smallest
$k_r$, or the largest wavelength.
For a cavity with radial extent $\delta r$ (set by the extent of the immiscible
region), $k_{r, \max} \simeq 2\pi / (\delta r)$\footnote{
This approximation is good for a thin cavity.
For larger cavities, the full $g$ mode frequency condition $\int
N\s{l(l+1)}^{1/2}/(\omega_g r)\;\mathrm{d}r = 2\pi n$ must be used, which yields
a logarithmic dependence on the cavity size.
}.
Then, $N^2$ also depends on $\delta r$ via the maximum extent of
compositional stratification possible in the rain region.
By expanding $N^2$ in terms of structural and thermodynamic derivatives, we find
that
\begin{align}
    \omega_g^2
        \simeq{}&
            \frac{l(l+1)}{4\pi^2}\p{\frac{\delta r}{r}}^2
            \times\s{
                -\frac{g}{\rho}
                \p{\pd{\rho}{S}}_{PY}
                \p{\rd{S}{r} - \p{\pd{S}{Y}}_{P\rho}\rd{Y}{r}}
            }.
\end{align}
But then, since the fluid satisfies marginal instability, some light algebra
shows that
\begin{align}
    0
        \approx{}&
            \rd{S}{r}
            - \s{
                \p{\pd{S}{Y}}_{PT}(1 - R_\rho)
                + \p{\pd{S}{Y}}_{P\rho}R_\rho
            }
            \rd{Y}{r},\\
    \frac{\omega_g^2}{\Omega_{\rm dyn}^2}
        \simeq{}&
            \Bigg\{\frac{g}{g_{\rm surf}}
               \p{\pd{\ln \rho}{S}}_{PY}
            \s{\p{\pd{S}{Y}}_{PT} - \p{\pd{S}{Y}}_{P\rho}}\Bigg\}
           (1 - R_\rho)
            \frac{l(l+1)}{4\pi^2}
            \p{\frac{R \,\delta r}{r^2}}
            \Delta Y,\label{eq:omegag}
\end{align}
Here, $g$ denotes the local gravitational acceleration, and $g_{\rm surf}$ that
at Saturn's surface.
The quantity in the curly braces is typically of order unity ($0.5$--$1$)
throughout the planet.
Thus, for an $\ell=2$ mode, the typical mode frequencies expected scale like
$\sim \sqrt{(1 - R_\rho)\Delta Y R (\delta r) / r^2}$.
As can be seen, this number generally remains below $1$, and does not reach
$\sim 2$ as would be required to be relevant to the W76.44 oscillation frequency.

\section{A Quantitative Condition for Stair Steps}\label{app:stairsteps}

In the main text, we remarked that our evolutions with Schwarzschild-like
convection ($R_\rho \to 0$) tend to be numerically unstable much more frequently
than our Ledoux-like convection ($R_\rho \to 1$).
This is due to the different evolution of so-called compositional ``stair
steps'' in the two limits (we specifically refer to those first identified by
\citealp{vazan2018jupiter}).
In this appendix, we reprise the discussion of \citet{vazan2018jupiter} more
quantitatively and record the challenges we incurred in evolving small-$R_\rho$
models.
We remark that these numerical challenges do not preclude all small-$R_\rho$
models (see the successful runs shown in \citealp{sur2025_rrho}), but they seem
to predominantly afflict models with stronger composition gradients, for reasons
that will soon be clear.

To build a conceptual model for convective mixing, we make a few simplifying
assumptions: (i) convection is the only process that transports energy and
composition (in reality, it dominates diffusive transport), and (ii) convective
regions are isentropic and homogeneous (closely satisfied in just a few
convective turnover times).
Furthermore, to simplify the discussion, take $Y'$ to be uniform, so that only
$Z$ dictates the composition profile.
Then, suppose the planet starts with an initial decreasing entropy profile
$S_0(m)$ and initial increasing composition profile $Z_0(m)$, such as that shown
in Fig.~\ref{fig:sat_rotnocore}.
As the surface cools, the surface convective layer expands downwards, such that
$m_{\rm stab}$ (the mass coordinate corresponding to $R_{\rm stab}$, the outer
extent of the stably stratified region) decreases.
Then, the entropy and composition must satisfy:
\begin{align}
    Z(m, t)
        &=
            \begin{cases}
                Z_0(m) & m < m_{\rm stab},\\
                \frac{1}{M - m_{\rm stab}}\int\limits_{m_{\rm stab}}^M
                    Z_0(m')\;\mathrm{d}m' & m > m_{\rm stab},
            \end{cases},\label{eq:app_Z_mt}\\
    S(m, t) &=
        \begin{cases}
            S_0(m) & m < m_{\rm stab},\\
            S_0(m_{\rm stab}) + \Delta S & m > m_{\rm stab}.
        \end{cases}\label{eq:app_S_mt},\\
    \Delta S
        &=
            \s{\p{\pd{S}{Z}}_{P\rho}R_\rho + \p{\pd{S}{Z}}_{PT}(1 - R_\rho)}
            \Delta Z,\label{eq:app_dS}
\end{align}
where $\Delta Z$ is the jump in composition across $m_{\rm stab}$ in
Eq.~\eqref{eq:app_Z_mt}.
Equation~\eqref{eq:app_Z_mt} corresponds to no composition evolution interior to
$m_{\rm stab}$ and the conservation of mass as the fluid exterior to $m_{\rm
stab}$ mixes.
Equation~\eqref{eq:app_S_mt} has the three required characteristics: primordial
entropy profile interior to $m_{\rm stab}$, isentropy exterior to $m_{\rm
stab}$, and marginal convective stability at $m_{\rm stab}$
(Eq.~\ref{eq:app_dS}).
Marginal convective stability is correct: if $\Delta S$ is too small (negative),
then convection will set in, and $m_{\rm stab}$ will propagate, while if $\Delta
S$ is too large, then $m_{\rm stab}$ should not yet have been reached.
Under the assumptions we have laid out, the evolution proceeds such that $m_{\rm
stab}$ proceeds inwards, successively driving more and more of the planet to be
convective as the surface of the planet cools \citep{knierim2024_erosion}.
Note that in standard convection (corresponding to the Ledoux condition
$R_{\rho} = 1$), the coefficient in Eq.~\eqref{eq:app_dS} is almost always
positive for the equations of state considered in our work.
Thus, $S_+(m_{\rm stab}, t) < S_-(m_{\rm stab}, t)$ when $m_{\rm stab}$ becomes
convective, and the extra energy released from the region just interior to
$m_{\rm stab}$ marginally stabilizes the surface convective region.

However, as $m_{\rm stab}$ decreases, $\Delta Z$ grows, since the integrated
average of $Z$ grows much more slowly than its local value at $m_{\rm stab}$.
This drives an increasingly steep temperature gradient across $m_{\rm stab}$.
In a realistic 1D evolutionary model, diffusive energy transport (conduction)
across the interface will then grow as $m_{\rm stab}$ propagates inwards.
By evaluating the temperature gradient across this interface under the condition
of marginal convective stability\footnote{
Most easily done by taking $v_{\rm conv}^2 \propto -\nabla + \nabla_{\rm ad} +
B$ using standard conventions \citep[e.g.][]{mankovich2021diffuse,
tejada2024_eos} and noting that $\nabla_{\rm ad}$ is to be
evaluated on either side of the interface only, is thus non-singular across the
interface, and so can be neglected.
},
we arrive at the result for the diffusive energy flux across the interface
\begin{equation}
    F_{\mathrm{diff}}
        = -\lambda_{\rm diff} R_\rho\frac{\Delta Z}{\Delta r}\p{\pd{T}{S}}_{PZ}
            \s{\p{\pd{S}{Z}}_{P\rho}
               - \p{\pd{S}{Z}}_{PT}},
\end{equation}
where $\lambda_{\rm diff}$ is typically the thermal conductivity in giant
planets.
We note that in numerical models of planets, $\Delta r$ is bounded from below by
the grid size, so $\Delta Z$ must attain some minimum threshold before the
diffusive flux is able to supply sufficient flux to transport energy across
$m_{\rm stab}$.
This picture is precisely that described qualitatively by
\citet{vazan2018jupiter} in their Section~3.2.
Indeed, we verify that \texttt{APPLE} satisfies continuity of total flux across
the stair steps in our models.

We first make a few comments on this result.
First, as described in \citet{vazan2018jupiter}, these stair steps are physical
in nature, despite their detailed properties being sensitive to the grid size.
Our description above provides a transparent description of this: the grid size
indeed sets the critical $\Delta Z$ at which $F_{\rm diff}$ will be sufficient
to transport energy across the interface, but the strengthening of diffusive
energy transport as $m_{\rm stab}$ decreases is a manifestly physical effect.
Second, in realistic planets, it is broadly suspected that such stair steps will
be erased by double-diffusive processes as well as convective overshoot
\citep{vazan2018jupiter}.
In our models, these stair steps do eventually disappear, but their
disappearance cannot be due to these processes, which are not handled in
\texttt{APPLE}.
Instead, we find that it seems to be both due to the changing
properties of the planet (the equation of state derivatives) resulting in an
inability to transport sufficiently large energy fluxes via diffusion and due to
violent structural changes each time a mass cell becomes convective, sometimes
driving the iterative structure solver away from the stair step structure.

Lastly, we comment on the relevance to the numerical stiffness of Schwarzschild
convection.
Throughout much of the planet, the coefficient in Eq.\eqref{eq:app_dS} is
negative when $R_\rho = 0$.
This requires that $S_+(m_{\rm stab}, t) > S_-(m_{\rm stab}, t)$ at marginal
stability.
However, since $S_-(m_{\rm stab}, t) = S_0(m_{\rm stab})$, this implies that
$S_+(m_{\rm stab}, t) > S_0(m_{\rm stab})$, i.e.\ the entropy of the base of the
convective envelope is \emph{higher} than it began.
This is difficult to attain via smooth evolution using the standard equations of
planet evolution.
This difficulty is not exclusive to our formation of the evolution equations in terms of the entropy: in terms of the temperature, there will always be a negative temperature gradient across the core-envelope boundary, which immediately be unstable under the Schwarzschild condition.

In our evolutions, the way such a profile is attained is that $S_-$
begins to decrease while $m_{\rm stab}$ is still some distance away,
such that the temperature remains continuous across the boundary.
In order to transport heat up to $m_{\rm stab}$, intermittent periods of convection set in that transport some material and heat from regions below
$m_{\rm stab}$.
Moreover, as this high-metallicity material propagates outwards, it tends to
enhance convection in its vicinity, as it is much hotter material from the
interior of the planet.
The combination of these two effects causes many of our models to grind to a
halt under our adaptive timestep, particularly when the advected material is
especially $Z$-rich.
This problem is significantly ameliorated as long as the chosen values of
$R_\rho$ are sufficiently large that Eq.~\eqref{eq:app_dS} is positive.

\end{document}